\documentclass{article}[11pt]

\usepackage{amsmath,amssymb,amsfonts}
\usepackage{physics}
\usepackage{dsfont}
\usepackage{graphicx}

\usepackage{tikz}

\usetikzlibrary{arrows,shapes,trees}

\newcommand{\ra}{\rightarrow}
\newcommand{\eps}{\epsilon}
\newcommand{\Hom}{{\rm Hom}}

\newcommand{\CC}{{\mathbb C}}
\newcommand{\ZZ}{{\mathbb Z}}
\newcommand{\RR}{{\mathbb R}}

\newcommand{\PP}{{\mathbb P}}

\newcommand{\tB}{{\tilde B}}

\newcommand{\cB}{\mathcal B}

\newcommand{\cF}{{\mathcal F}}

\newcommand{\cC}{\mathcal C}

\newcommand{\s}{{\mathfrak s}}
\newcommand{\Eta}{\zeta}

\newcommand{\zp}{{\mathcal Z}}
\newcommand{\dw}{{\mathcal D}}
\newcommand{\athree}{{\mathcal J}}

\usepackage{hyperref}
\newcommand{\footremember}[2]{%
    \footnote{#2}
    \newcounter{#1}
    \setcounter{#1}{\value{footnote}}%
}

\title{Fermionic SPT phases in higher dimensions and bosonization}
\author{Anton Kapustin\footremember{anton}{Walter Burke Institute for Theoretical Physics, California Institute of Technology}, Ryan Thorngren\footremember{ryan}{Department of Mathematics, University of Calfornia, Berkeley}}

\begin{document}

\maketitle

\begin{abstract}
We discuss bosonization and Fermionic Short-Range-Entangled (FSRE) phases of matter in one, two, and three spatial dimensions, emphasizing the physical meaning of the cohomological parameters which label such phases and the connection with higher-form symmetries. We propose a classification scheme for fermionic SPT phases in three spatial dimensions with an arbitrary finite point symmetry $G$. It generalizes the supercohomology of Gu and Wen. We argue that the most general such phase can be obtained from a bosonic ``shadow'' by condensing both fermionic particles and strings. 
\end{abstract}

\section{Introduction and summary}

A  topological phase of matter with a symmetry $G$ is an equivalence class of gapped quantum lattice systems with a symmetry $G$. One can study either ground states or Hamiltonians. For classification purposes, it is the same\cite{KitaevTalk}. In terms of ground states, the equivalence relations are of two kinds: tensoring with a product state, eg. the ground state of a trivial paramagnet (this adds new degrees of freedom), and local unitary transformations of the ground state commuting with $G$. Topological phases of matter can be ``stacked together'', by taking tensor product of Hilbert spaces, Hamiltonians, and ground states, and taking the $G$ symmetry of the stack to be the diagonal one. This operation makes the set of topological phases with a symmetry $G$ into a commutative unital semigroup, a set with an associative and commutative binary operation and a neutral element, but not necessarily with an inverse for every element.  A short-range-entangled (SRE) topological phase with symmetry $G$ is a topological phase with symmetry $G$ which has an inverse. SRE topological phases in $d$ spatial dimensions with symmetry $G$ form an abelian group.

SRE topological phases are interesting in part because they are more manageable than general topological phases but still retain many interesting topological properties. Fermionic SRE topological phases (FSRE phases) are particularly rich.  Free FSRE phases, ie. equivalence classes of quadratic Hamiltonians of hopping fermions, have been classified in all spatial dimensions \cite{Kitaevtable,Ryuetal}. In the interacting case, there is a fairly complete picture of FSRE phases in dimensions $1$ and $2$ \cite{ChenGuWen,FidKit,GK,BGK} (the abelian group structure on the set of 1d FSREs was recently studied in \cite{Bultincketal,KTY2}). Gu and Wen also constructed a large class of FSRE phases in all dimensions using the ``supercohomology'' approach \cite{GuWen}. But it is clear by now that this construction does not produce all possible FSRE phases.

It was conjectured in \cite{Kapustinetal} that FSRE phases can be classified using spin-cobordism\footnote{This is a refinement of the supercohomology proposal of \cite{GuWen}.} of the classifying space of $G$. This conjecture is supported by a recent mathematical result that relates (spin) cobordisms with unitary invertible (spin) TQFT \cite{FreedHopkins}. The drawback of this approach is that the relation between TQFTs and topological phases of matter is not well understood. In particular, given a spin-cobordism class it is not clear in general how to construct a lattice fermionic system which belongs to the corresponding FSRE phase. Neither is it clear which physical properties distinguish systems corresponding to different cobordism classes.   

In dimensions $1$ and $2$ these problems have been solved, at least in the case when $G$ acts unitarily. For example, 1d FSRE phases are classified by a triple \cite{ChenGuWen,FidKit}:
\begin{equation}\label{onedFSRE}
(\nu,\rho,\sigma)\in H^2(G,\RR/\ZZ)\times H^1(G,\ZZ_2)\times H^0(G,\ZZ_2)
\end{equation}
To each such triple one can assign a concrete integrable lattice Hamiltonian  as well as a spin-cobordism class of $BG$ \cite{GK}. The physical significance of each member of the triple is understood (they describe properties of the edge  modes of the system). Similar results for 2d FSRE systems have been obtained in \cite{GK,BGK}.

The main goal of this paper is to extend some of these results to dimension $3$. Our approach is based on the idea of bosonization/fermionization. It is a well-known result that every lattice fermionic system in one spatial dimension corresponds to a lattice bosonic system with a global  $\ZZ_2$ symmetry. This is usually explained using the Jordan-Wigner transformation. In \cite{GK} it was argued that one can obtain fermionic systems in $d$ spatial dimensions starting from bosonic systems with a global $(d-1)$-form $\ZZ_2$ symmetry generated by a fermionic quasiparticle. More precisely, the $\ZZ_2$ symmetry must have a particular 't Hooft anomaly which is trivialized when the spin structure is introduced. The fermionic system can be recovered by gauging the $\ZZ_2$ symmetry, i.e. by coupling the bosonic system to a dynamical $d$-form gauge field valued in $\ZZ_2$ as well as a simple fermionic system. In 1d and 2d every FSRE phase arises in this way from a suitable bosonic system, and it is natural to conjecture that this is also true in higher dimensions. 

In fact, we will argue that in 3d a new phenomenon occurs which makes the bosonization approach a bit more involved. Namely, the fermion parity operator $(-1)^F$  can get contributions from both particle and string states, and the string contribution cannot be written in a local way. Microscopically, these strings carry a 1d FSRE phase (the Kitaev chain \cite{Kitaevtable}), which may have a fermionic ground state depending on how it is embedded into space. We call these objects Kitaev strings. From the mathematical viewpoint, this means that the bosonic shadow has both 2-form and 1-form global $\ZZ_2$ symmetries, with a nontrivial ``interaction'' between them, and both need to be gauged in order to get an FSRE phases. We propose a generalization of the Gu-Wen supercohomology which accounts for this new phenomenon. We also write down a concrete 3d lattice bosonic model which, when coupled to a background $G$ gauge field, gives the bosonic shadow of a general 3d FSRE phase. This theory is interesting in its own right as a very simple non-abelian 3+1d topological order, analogous to the Ising anyons in 2+1d.

The organization of the paper is as follows. In sections 2 and 3 we recall topological bosonization in one and two spatial dimensions and how it is used to classify FSRE phases. We also interpret the classification in terms of properties of domain walls and their junctions in a broken symmetry phase. In section 4 we describe our proposal for 3d bosonization and propose a classification of 3d FSRE phases. In section 5 we write down a 3d bosonic model which can serve as a bosonic shadow for 3d FSRE phases. In section 6 we briefly discuss a new class of 3d phases which seem to be neither bosonic nor fermionic, although they contain ``fermionic strings''. In section 7 we summarize our results and discuss possible higher-dimensional generalizations.

We will be interested in models where the fermion number is conserved modulo $2$. Accordingly, the eigenvalues of the fermion number operator $F$ are defined only as elements of $\ZZ_2=\ZZ/2\ZZ$. The eigenvalues of the fermion parity operator $(-1)^F$ are $\pm 1$. We will freely use simplicial cochains and operations on them, including Steenrod squares. Some properties of Steenrod squares and Stiefel-Whitney classes are recalled in Appendix \ref{steenrodreview}.

A. K. would like to thank Greg Brumfiel, John Morgan and Anibal Medina for communicating to him some of their unpublished results. R. T. would like to thank Dominic Williamson, Dave Aasen, and Ethan Lake for many enlightening discussions. This paper was supported in part by the U.S. Department of Energy, Office of Science, Office of High Energy Physics, under Award Number DE-SC0011632. The work of A. K. was also supported by the Simons Investigator Award. R. T. is supported by an NSF GRFP grant. A. K. and R. T. are grateful to KITP, Santa Barbara, for hospitality during the initial stages of this project.

\section{Bosonization and FSRE phases in one spatial dimension}

\subsection{Bosonization in 1d}

It is well-known that 1d fermionic systems can be mapped to bosonic systems with $\ZZ_2$ symmetry by means of the Jordan-Wigner transformation\cite{Haldane}. This tranformation is not an equivalence, as it does not preserve certain physical properties. For example, it maps the Majorana chain \cite{FidKit} (a discretization of the  massive Majorana fermion) to the quantum Ising chain. Depending on the values of the parameters, the latter model can have a doubly-degenerate ground state. On the other hand, the ground state of the Majorana chain is always unique. It is best to think about the JW transformation as ``gauging the fermion parity". This becomes more obvious when one considers the bosonization transformation on a circle\cite{Bentov}. While the massive fermion on a circle requires a spin structure, the corresponding bosonic system does not. On the other hand, since it has a $\ZZ_2$ symmetry, it can be coupled to a $\ZZ_2$ gauge field. To obtain the bosonic Hilbert space from the fermionic Hilbert space one has to ``sum over spin structures". Conversely, the fermionic Hilbert space can be obtained from the bosonic one by ``summing over $\ZZ_2$ gauge fields.'' The scare quotes indicate that certain topological terms are important in these sums.

One can describe the connection between bosonic and fermionic Hilbert spaces on a circle in complete generality. The bosonic Hilbert space has a $\ZZ_2$-untwisted sector and a $\ZZ_2$-twisted sector, which we denote $\cB_0$ and $\cB_1$. 
Each of these can be further decomposed into eigenspaces of the $\ZZ_2$ global symmetry:
$$
\cB_0=\cB_0^+\oplus\cB_0^-,\quad \cB_1=\cB_1^+\oplus\cB_1^-.
$$ 
On the other hand, the fermionic Hilbert space has an NS sector and a R sector, which we denote $\cF_{NS}$ and $\cF_R$, and each of them decomposes into eigenspaces of the fermion parity $P$:
$$
\cF_{NS}=\cF_{NS}^+\oplus\cF_{NS}^-,\quad \cF_R=\cF_R^+\oplus\cF_R^-.
$$
These decompositions are related as follows:\footnote{There is an ambiguity here, since we can tensor an arbitrary fermionic phase with a nontrivial fermionic SRE phase (the negative-mass Majorana chain \cite{Kitaevtable}) and thereby flip the fermion parity of the Ramond-sector states while leaving the NS sector unaffected. This amounts to multiplying by the Arf invariant when we sum over spin structures and reverses the correspondence between fermionic and bosonic phases. We choose our conventions so that higher-dimensional generalizations are more straightforward.}
$$
\cF_{NS}^+=\cB_0^+,\quad \cF_{NS}^-=\cB_1^-,\quad \cF_R^+=\cB_0^-,\quad \cF_R^-=\cB_1^+.
$$
In particular, the $(-1)^F=1$ component of the total fermionic Hilbert space, $\cF_{NS}^+\oplus \cF_R^+$, is the untwisted sector of the bosonic theory $\cB_0^+\oplus \cB_0^-$, while the $(-1)^F=-1$ component $\cF_{NS}^-\oplus \cF_R^-$, is the twisted sector of the bosonic theory $\cB_1^-\oplus \cB_1^+$. 

These relations can be interpreted as follows: to get the fermionic Hilbert space from the bosonic one, one gauges the $\ZZ_2$ symmetry and identifies the holonomy of the $\ZZ_2$ gauge field $(-1)^{\alpha}$ as the fermion parity $P$. For each value of $\alpha\in\ZZ_2$, one needs to  project to a particular value of the $\ZZ_2$-charge to select either the NS or R sector states: More precisely, if we label the spin structures by $\s\in\ZZ_2$ so that $\s=0$ corresponds to the NS sector and $\s=1$ corresponds to the R sector, then the generator of $\ZZ_2$ acts in the sector with the holonomy $(-1)^{\alpha}$ with the weight $(-1)^{\s+\alpha}$. Note that the weight is a (exp-)linear function both of the spin structure and the $\ZZ_2$ gauge field on a circle. 

As an example, consider the Majorana chain \cite{Kitaevtable,FidKit} and the quantum Ising chain.
The quantum Ising chain has a gapped phase with an unbroken $\ZZ_2$ (paramagnet) and a gapped phase with a spontaneously broken $\ZZ_2$ (ferromagnet). Consider the limit of an infinite energy gap. Then in the unbroken phase the system has a unique ground state both for the trivial and the nontrivial $\ZZ_2$ gauge field. On the other hand, in the broken phase, the system has two ground states with a trivial $\ZZ_2$ gauge field (a $\ZZ$-even one and a $\ZZ_2$-odd one), and no ground states when the $\ZZ_2$ gauge field is turned on because in the limit of infinite energy gap the energy of the domain wall between the two vacua is infinite. The Majorana chain also has two phases, depending on the sign of the parameter which corresponds to the fermion mass in the continuum limit. For both signs of the mass, there is a unique ground state for either choice of the spin structure on a circle. The difference is that for a positive mass the Ramond-sector ground state has $(-1)^F=1$, while for a negative mass it has $(-1)^F=-1$. The ground state in the NS sector has $(-1)^F=1$ in both cases. 

The JW transformation maps the positive-mass Majorana chain to the Ising chain with a spontaneously broken $\ZZ_2$, while the negative-mass Majorana chain is mapped to the Ising chain with an unbroken $\ZZ_2$. 

Note that the Majorana chain (for either sign of the mass) is an FSRE, but the quantum Ising chain in a phase with a spontaneously broken $\ZZ_2$ is not a bosonic SRE phase. Thus bosonization and fermionization do not map SRE phases to SRE phases. This also applies in higher dimensions, as we will see.

When considering 1d systems on a circle, it is easy to mistake a spin structure for a $\ZZ_2$ gauge field. The distinction between them becomes clearer when we consider systems on a curved space-time with a nontrivial topology. It will be useful to write down a relation between the partition functions of the fermionic theory and its bosonic ``shadow'' on a general Riemann surface $M$. The fermionic partition function depends on a spin structure on $M$, while the bosonic partition function depends on a $\ZZ_2$ gauge field $\alpha$ (i.e. an element of $H^1(M,\ZZ_2)$). 

A nice way to think about a spin structure on $M$ is as follows \cite{Atiyah}: every spin structure $\eta$ gives rise to a quadratic function $q _\eta: H^1(M,\ZZ_2)\ra\ZZ_2$, such that 
\begin{equation}\label{quadratic2d}
q_\eta(a+b)-q_\eta(a)-q_\eta(b)=\int_M a\cup b. 
\end{equation}
Conversely, every such quadratic function corresponds to a spin structure on $M$. One says that $q_\eta$ is a quadratic refinement of the bilinear form $(a,b)\mapsto \int_M a\cup b$. Note that the set of spin structures is not an abelian group (there is no natural way to define a group operation on the set of quadratic refinements of a fixed bilinear form). On the other hand, the set of equivalence classes of $\ZZ_2$ gauge fields is an abelian group. We note for future use that the latter group naturally acts on the set of spin structures: for all $\alpha,\alpha'\in H^1(M,\ZZ_2)$ we let
$$
q_{\eta+\alpha}(\alpha')=q_\eta(\alpha')+\int_M \alpha\cup\alpha'.
$$

Given this relation between spin structures and quadratic refinements, the relation between the partition functions can be written as a nonlinear discrete Fourier transform:
$$
Z_f(\eta)=\frac{1}{2^{b_1(M)/2}}\sum_{\alpha\in H^1(M,\ZZ_2)} Z_b(\alpha) (-1)^{q_\eta(\alpha)}.
$$

In our example of the Ising/Majorana correspondence, $Z_b(\alpha)$ is a delta function $\delta(\alpha)$ setting $\alpha = 0$ in the ferromagnetic phase (because of the infinite energy of the domain wall) and the constant 1 in the paramagnetic phase. So the former coincides with the constant fermionic partition function while the latter coincides with the Arf invariant, which is the partition function of the Kitaev chain \cite{Kapustinetal}, agreeing with what we expect from the microscopic JW transformation.

\subsection{FSRE phases in 1d}

FSREs in 1d (with arbitrary interactions) have been classified in \cite{ChenGuWen,FidKit} using bosonization and Matrix Product States. See also \cite{Bultincketal,KTY2}, where the same results were obtained using fermionic MPS. The result is that the set of FSRE phases with a unitary symmetry $G$ is classified\footnote{For simplicity, we are assuming that the total symmetry is $G$ times fermion parity, rather than an extension of $G$ by fermion parity. The generalization to nontrivial extensions is straightforward.}  by triples $(\sigma,\rho,\nu)$ of group cohomology classes (\ref{onedFSRE}). All of these parameters can be be interpreted in terms of properties of the edge zero modes. The parameter $\sigma\in H^0(G,\ZZ_2)=\ZZ_2$ is the number modulo two of Majorana zero modes at each edge of the system. For example, the negative-mass Majorana chain \cite{Kitaevtable} has $\sigma=1$ and a single Majorana zero mode at every edge. The parameter $\rho\in H^1(G,\ZZ_2)$ tells us whether a particular element $g\in G$ commutes ($\rho(g)=0$) or anti-commutes ($\rho(g)=1$) with the fermion parity $(-1)^F$ when acting on the edge zero modes. The parameter $\nu\in H^2(G,\RR/\ZZ)$ controls the projective nature of the action of $G$ on the edge zero modes. If any one of these parameters is non-vanishing, the system must have nontrivial edge zero modes, and therefore the ground state on an interval is degenerate in the large-volume limit. 

It is instructive, although somewhat nontrivial, to interpret these parameters without appealing to the edge zero modes \cite{KTY2}. It is helpful to introduce a nontrivial spacetime geometry and a fixed background $G$ gauge field. Let us imagine that the IR limit of the system is described by a unitary continuum 2d quantum field theory, then we can Wick-rotate it and place it on an arbitrary Riemann surface $\Sigma$, perhaps with a nonempty boundary $\partial\Sigma$. For a fermionic system, this requires choosing a spin structure on $\Sigma$, which also induces a spin structure on each boundary circle in $\partial\Sigma$.
There are two spin structures on a circle: periodic (Ramond) and anti-periodic (Neveu-Schwarz). They are also known as non-bounding and bounding spin structures, respectively, since the NS spin structure on a circle can be obtained by restricting the unique spin structure on a disk, while the Ramond  spin structure cannot be so obtained. Since we are dealing with an FSRE phase, the ground state on a circle is non-degenerate for either choice of the spin structure, and one can show that in the NS sector it is always parity-even \cite{MooreSegal}. The parameter $\sigma$ tells us whether the ground state in the Ramond sector is bosonic ($\sigma=0$) or fermionic ($\sigma=1$). 

One can also couple the system to a flat $G$ gauge field and consider the ground states on a circle with a holonomy $g\in G$ (and an arbitrary spin structure). For any $g\in G$ there is a unique ground state (again by the SRE assumption). The parameter $\rho(g)$ tells us whether it is bosonic a fermionic for the NS spin structure (for the Ramond spin structure, the fermion parity of the ground state is shifted by $\sigma$). When the symmetry $G$ is broken, turning on a holonomy $g$ around the circle leads to a particle-like domain wall; the paremeter $\rho(g)$ tells us whether it is bosonic or fermionic.

Finally, the parameter $\nu$ describes the ``S-matrix'' of the domain walls obtained when the symmetry $G$ is spontaneously broken. To be more precise, let us assume that $\rho=\sigma=0$. Then all domain walls are bosonic, and since the theory is trivial away from the domain walls, one should be able to compute the partition function by summing over possible domain-wall worldlines. The parameter $\nu(g_1,g_2)$ is a phase attached to a junction of domain walls labeled by $g_1$ and $g_2$. 

Together, these parameters define a 2-dimensional spin cobordism class of $BG$ via the Atiyah-Hirzebruch spectral sequence.

\section{Bosonization and FSRE phases in two spatial dimensions}

\subsection{Bosonization in 2d}

Recently, it has been shown that a 2+1d lattice fermionic system can be obtained from a 2+1d bosonic system (its bosonic ``shadow'') with an anomalous $\ZZ_2$ 1-form symmetry. Let us remind what this means \cite{gensym}. A parameter of a global 1-form $\ZZ_2$ symmetry is a $\ZZ_2$ gauge field, i.e. a 1-cocycle (Cech or simplicial) with values in $\ZZ_2$, defined up to a $\ZZ_2$ gauge transformation (i.e. up to adding an exact 1-cocycle). This symmetry is assumed to preserve the action, but cannot be gauged. That is, one cannot promote the parameter $\lambda$ to a general $\ZZ_2$-valued 1-cochain even at the expense of introducing a 2-form gauge field $B$ (i.e. a 2-cocycle with values in $\ZZ_2$ which transforms as $B\ra B+\delta\lambda$) while maintaining gauge invariance. The anomaly of a bosonic shadow has a very specific form: the partition function $Z_b(B)$ on a closed oriented 3-manifold $Y$ transforms under $B\ra B+\delta\lambda$  by a factor
\begin{equation}\label{anomaly3d}
(-1)^{\int_Y \left(\lambda\cup B+B\cup\lambda+\lambda\cup \delta\lambda\right)}.
\end{equation}
It was shown in \cite{GK,BGK} that one can obtain the fermionic partition function by performing a nonlinear discrete Fourier transform:
\begin{equation}\label{Fourier3d}
Z_f(\Eta)\sim \sum_{[B]\in H^2(Y,\ZZ_2)} Z_b(B) (-1)^{Q_\Eta(B)}
\end{equation}
Here we use an observation \cite{GK} that to every spin structure $\Eta$ on a triangulated closed oriented 3-manifold $Y$ one can associate a quadratic function $Q_\Eta: Z^2(Y,\ZZ_2)\ra \ZZ_2$ which under $B\ra B+\delta\lambda$ transforms as
\begin{equation}\label{quadratictransf3d}
Q_\Eta(B+\delta\lambda)=Q_\Eta(B)+\int_Y \left(\lambda\cup B+B\cup\lambda+\lambda\cup \delta\lambda\right)
\end{equation}
The construction and properties of the function $Q_\Eta$ are discussed in Appendix \ref{QEta}. Thanks to (\ref{quadratictransf3d}), the summand in (\ref{Fourier3d}) is a well-defined function on $H^2(Y,\ZZ_2)$. 

Unlike in 2d, the definition of the quadratic function $Q_\Eta$ depends on additional choices:  a branching structure on the triangulation. The bilinear form on $Z^2(Y,\ZZ_2)$ corresponding to the quadratic function $Q_\Eta$ is independent of $\Eta$ but depends on these extra choices:
\begin{equation}\label{quadratic3d}
Q_\Eta(B+B')-Q_\Eta(B)-Q_\Eta(B')=\int_Y B\cup_1 B',
\end{equation}
where $\cup_1$ is a certain bilinear operation $C^2(Y,\ZZ_2) \times C^2(Y,\ZZ_2)\ra C^3(Y,\ZZ_2)$ introduced by Steenrod \cite{Steenrod_higher} (see Appendix \ref{steenrodreview}). One can show that spin structures on $Y$ are in one-to-one correspondence with quadratic refinements of this bilinear form which transform according to (\ref{quadratictransf3d}), see Appendix \ref{QEta} and\cite{MorganBrumfiel}. 

The equation (\ref{Fourier3d}) says that the fermionic theory is obtained from the bosonic one by gauging the $\ZZ_2$ 1-form symmetry. The factor $(-1)^{Q_\Eta(B)}$ is needed to cancel the gauge anomaly. It is instructive to see how the gauging works on the Hamiltonian level. Consider a space-time of the form $Y=M\times\RR$, where $M$ is a closed Riemann surface. There are two sectors in the gauged bosonic theory distinguished by the flux of the 2-form gauge field $B$ through $M$. The untwisted sector $\int_M B=0$ is identified with the $(-1)^F=1$ sector of the fermionic Hilbert space, while the twisted sector $\int_M B=1$ is identified with the $(-1)^F=-1$ sector of the fermionic Hilbert space. The gauge 1-form $\ZZ_2$ symmetry acts in each sector by unitary operators $U_\lambda$, $\lambda\in C^1(M,\ZZ_2)$. By fixing a gauge, we can assume that $\lambda$ is closed, so that each sector is acted upon by $Z^1(M,\ZZ_2)$. This action is projective because of the 't Hooft anomaly. The corresponding 2-cocycle is computed following a standard procedure, see Appendix \ref{descent} and \cite{BGK}. We get
\begin{equation}\label{projcocycle3d}
U_\lambda U_{\lambda'}=(-1)^{\int_M \lambda\cup\lambda'} U_{\lambda+\lambda'}.
\end{equation}
In particular $U_\lambda^2=1$. As in the 1d case, the sector corresponding to a particular spin structure $\eta$ on $M$ is  obtained by decomposing the Hilbert space into eigenspaces of $U_\lambda$, namely
$$
U_\lambda \vert\Psi,\eta\rangle=(-1)^{q_\eta(\lambda)}\vert \Psi,\eta\rangle
$$
This is consistent with (\ref{projcocycle3d}) thanks to (\ref{quadratic2d}).\footnote{Alternatively, one can say that the cocycle in (\ref{projcocycle3d}) can be trivialized by defining $\tilde U_\lambda=(-1)^{q_\eta(\lambda)} U_\lambda$ and requiring physical states to be invariant under $\tilde U_\lambda$ for all $\lambda\in Z^1(M,\ZZ_2)$.}

\subsection{FSRE phases in 2d}

Let us now recall the classification of 2d FSRE phases proposed in \cite{BGK}. They are labeled by triples 
$$
(\nu,\rho,\sigma)\in C^2(BG,\RR/\ZZ)\times Z^2(BG,\ZZ_2)\times Z^1(BG,\ZZ_2),
$$
which satisfy the equations
$$
\delta\nu=\frac12 \rho\cup\rho,\quad \delta\rho=0,\quad \delta\sigma=0.
$$
The first two of these are the Gu-Wen equations which describe supercohomology phases. The bosonic shadow of all these FSRE phases can be taken to be the toric code equivariantized with respect to $G$ \cite{BGK}. In particular, the homomorphism $\sigma:G\ra\ZZ_2$ tells us which elements of $G$ exchange the $e$ and $m$ excitations of the toric code. The toric code has an action
$$
\frac12 \int_Y b da
$$
and a global  1-form $\ZZ_2$ symmetry which acts by $a\mapsto a+\lambda,$ $b\mapsto b+\lambda$, $\lambda\in Z^1(Y,\ZZ_2)$. One can check that this 1-form symmetry has the right 't Hooft anomaly. 

One can interpret the data $(\nu,\rho,\sigma)$ in physical terms. As stated above, a nonzero $\sigma(g)$ means that the element $g$ acts as particle-vortex symmetry of the toric code. This implies that an insertion of a flux $g$ of the background gauge field carries a Majorana zero mode (or more precisely, an odd number of Majorana zero modes). In the symmetry-broken phase, this insertion becomes an endpoint of a domain wall, and thus the corresponding domain wall carries a negative-mass Majorana chain. In what follows we will call a 1d defect with this property a Kitaev string. Let us denote by $\dw_g$ the domain wall corresponding to the group element $g$. Note that since fusing $\dw_g$ and $\dw_h$ produces $\dw_{gh}$, and the number of Majorana zero modes must be preserved modulo $2$, we must have 
\begin{equation}\label{sigmaishomomorphism}
\sigma(gh)=\sigma(g)+\sigma(h),
\end{equation}
i.e. $\sigma$ is a homomorphism. 

The parameter $\rho(g,h)\in Z^2(G,\ZZ_2)$ is most easily interpreted if  $\sigma=0$. Then the endpoint of each domain wall carries no fermionic zero modes, and one might as well assume that the endpoint has fermion parity zero. But when considering networks of domain walls, we might need to assign fermion parity $\rho(g,h)\in\ZZ_2$ to each triple junction, where $\dw_g, \dw_h,$ and $\dw_{h^{-1}g^{-1}}$ meet. Requiring that the fermion number of the network does not change under Pachner moves, one gets a constraint saying that $\rho(g,h)$ is a 2-cocycle (with values in $\ZZ_2$). Equivalently, one may consider the surface of a tetrahedron, and regard each edge as a domain wall. Since this network can be consistently continued into the interior of the tetrahedron, the fermion parity of the network must vanish. This again gives the condition $\delta\rho=0$.

Note also that since every domain wall has two ends, we can shift the fermion parity of the endpoint of $\dw_g$  by $f(g)\in\ZZ_2$ without changing the net fermion parity of the network. This shifts $\rho(g,h)$ by a coboundary:
$$
\rho(g,h)\mapsto \rho(g,h)+f(g)+f(h)+f(gh).
$$
Thus only the cohomology class of $\rho$ has a physical meaning.

When $\sigma(g)$ is non-vanishing, the situation is not very different. The key point is that at the junction of three domain walls we have an even number of Majorana zero modes, thanks to the condition (\ref{sigmaishomomorphism}). They act irreducibly on a fermionic Fock space, and one can imagine turning on a local interaction at the junction that lifts the degeneracy and makes one of these states the ground state. The fermion parity of this ground state is $\rho(g,h)$. The same arguments as above show that $\rho(g,h)$ is a 2-cocycle defined up to a coboundary.

The parameter $\nu(g,h,k)\in\RR/\ZZ$ has the same meaning as in the bosonic case, i.e. it describes the amplitude assigned to a point-like junction of four domain wall worldsheets in space-time. To derive a constraint on it, one needs to consider a 3-sphere triangulated into a union of four tetrahedra, pass to the dual cell complex and insert a domain wall along every 2-face of this cell complex. On the one hand, the amplitude muct be trivial, because such a configuration of domain walls can be created out of a trivial one. On the other hand, one can evaluate it taking into account the fermionic statistics of the triple domain wall junctions \cite{BGK}. The resulting constraint is the Gu-Wen equation
$$
\delta\nu=\frac12\rho\cup\rho. 
$$

When the parameter $\sigma$ is nontrivial, some domain walls are Kitaev strings  and consequently carry fermion number when wrapping cycles with Ramond spin structure. Note that a homologically trivial Kitaev string automatically carries zero fermion number, because the spin structure induced on it by the spin structure in the ambient space is of the NS type. Therefore the contribution of the Kitaev strings to the fermion number is nonlocal and depends on the homology class of the string network. To determine its form, note first that the homology class of the Kitaev strings is the Poicare-dual of $\sigma(A)\in H^1(M,\ZZ_2)$, where $A$ is the $G$ gauge field on $M$. Assuming that the fermion number depends only on the homology class of the string, we may assume that the Kitaev string wraps a closed curve $\gamma$ on $M$ whose homology class is dual to $\sigma(A)$. Then the spin structure induced on $\gamma$ is Ramond precisely if $q_\eta(\sigma(A))=1$. Therefore we can identify $q_\eta(\sigma(A))$ with the contribution of Kitaev strings to the fermion number $F$. Note that it is nonlocal, as expected, and conserved. This explains why we could ignore it when identifying the fermion number with a local expresson  $\int_M B$: in 2d FSRE phases, the particle and strings contrubutions to the fermion number are separately conserved.

As in the 1+1D case, these triples define spin cobordism classes of $BG$ via the Atiyah-Hirzebruch spectral sequence.

\subsection{The string-net ground state}

In this section we discuss ground states of a simple lattice model which is a bosonic shadow of the trivial 2d FSRE phase following \cite{BGK}. This is a warm-up for a similar discussion of 3d FSRE phases in later sections. We need a bosonic TQFT which has a 1-form $\ZZ_2$ symmetry with the correct anomaly. A bosonic TQFT can be constructed from a spherical fusion category $\cC$. Its objects can be thought of as boundary line defects for a particular boundary condition. Bulk line defects are described by objects in a modular tensor category $Z(\cC)$, the Drinfeld center of $\cC$.  A generator of a 1-form $\ZZ_2$ symmetry is a bulk line defect and thus corresponds to an object $\psi\in Z(\cC)$ with a fusion rule $\psi\circ\psi\simeq 1$. Such an object has topological spin $\theta_\psi$ which satisfies $\theta_\psi^4=1$. It measures the anomaly of the 1-form $\ZZ_2$ symmetry. Since we want the anomaly to be of order $2$, the topological spin must be $-1$, i.e. $\psi$ must be a fermion. 

The simplest 2+1d TQFT with these properties is the $\ZZ_2$ gauge theory, also known as the toric code. The corresponding category $\cC$ is the category of $\ZZ_2$ graded vector spaces and has two irreducible objects: $1$ and $F$, with the fusion rule $F\circ F\simeq 1$. One can think of the boundary line defect  $F$ as the result of fusing $\psi$ with the boundary. The toric code has two more irreducible line defects, $e$ and $m$, such that $e\circ m\simeq\psi$, and $e\circ e\simeq m\circ m\simeq 1$. The objects $e$ and $m$ are bosons ($\theta_e=\theta_m=1$) and thus correspond to non-anomalous $\ZZ_2$ symmetries, but since they are muutually nonlocal, their bound state $\psi$ is a fermion. 

Let the spatial slice be a closed oriented 2d manifold $M$ with a chosen triangulation. The toric code has  $\vert H^1(M,\ZZ_2)\vert$ linearly independent ground states on $M$. 
The string-net construction describes these ground states as particular linear combinations of states $\vert \gamma\rangle$, where $\gamma\in Z^1(M,\ZZ_2)$. A 1-cocycle on a triangulated surface can be thought of more geometrically as a 1-cycle on a dual cell complex, ie. a bunch of closed curves, a ``string net". The string-net Hamiltonian is a commuting projector Hamiltonian whose ground states have the property that the coefficient $C(\gamma)$ of the state $\vert\gamma\rangle$ is invariant under local rearrangements of the string-net which do not change its homology class, or dually, the cohomology class $[\gamma]\in H^1(M,\ZZ_2)$. Thus a general ground state is
$$
\sum_{\gamma\in Z^1(M,\ZZ_2)} C([\gamma])\vert\gamma\rangle.
$$

There are several natural bases in the space of ground states associated to various 1-form $\ZZ_2$ symmetries of the toric code. The most obvious basis
$$
\vert\Psi,[\beta]\rangle=\sum_{[\gamma]=[\beta]}\vert\gamma\rangle,\quad [\beta]\in H^1(M,\ZZ_2)
$$
can be characterized by the property that $\vert\Psi,[\beta]\rangle$ is a simultaneous eigenvector of the 1-form symmetry transformations 
$$
\vert\gamma\rangle\mapsto (-1)^{\int_M \alpha\cup\gamma} \vert\gamma\rangle,\quad \alpha\in Z^1(M,\ZZ_2).
$$
The 1-form symmetry which acts by
$$
\vert\gamma\rangle\mapsto \vert\gamma+\alpha\rangle,\quad \alpha\in Z^1(M,\ZZ_2)
$$
has simultaneous eigenvectors of the form
$$
\vert\Psi',[\beta]\rangle=\sum_\gamma (-1)^{\int_M \beta\cup\gamma} \vert\gamma\rangle.
$$
These two 1-form symmetries are non-anomalous and correspond to $e$ and $m$ bulk line defects.

The ``diagonal'' 1-form symmetry which acts by
\begin{equation}\label{anomalousoneform}
\vert\gamma\rangle\mapsto (-1)^{\int_M \alpha\cup\gamma} \vert\gamma+\alpha\rangle,\quad \alpha\in Z^1(M,\ZZ_2)
\end{equation}
corresponds to the bulk line defect $\psi$. Its simultaneous eigenvectors are labeled by spin structures $\eta$:
$$
\vert\Psi'',\eta\rangle=\sum_\gamma (-1)^{q_\eta(\gamma)} \vert\gamma\rangle.
$$
The anomalous 1-form symmetry (\ref{anomalousoneform}) acts on these states as follows:
$$
\alpha: \vert\Psi'',\eta\rangle \mapsto (-1)^{q_\eta(\alpha)} \vert\Psi'',\eta\rangle.
$$
Upon gauging the 1-form $\ZZ_2$ symmetry, the state $\vert\Psi'',\eta\rangle$ gives rise to the unique ground state of the fermionic TQFT on the spin manifold $(M,\eta)$. In order to get a nontrivial FSRE with symmetry $G$, one has to couple the toric code to a background $G$ gauge field. As explained in \cite{BGK}, this leads to the most general 2d FSRE with parameters $(\nu,\rho,\sigma)$.

\section{Bosonization and FSRE phases in three spatial dimensions}

\subsection{Bosonization in 3d}

It was mentioned in \cite{GK} that one should be able to construct fermionic phases in 3d from bosonic phases with an anomalous global 2-form $\ZZ_2$ symmetry. The anomaly is again quite special: it trivializes when a spin structure is specified.

The most concise way to describe the anomaly is to write down a 5d topological action for a 3-form $\ZZ_2$ gauge field $C\in Z^3(P,\ZZ_2)$ whose variation is a  boundary term cancelling the variation of the partition function of the anomalous theory on $\partial P=X$. In the present case, this anomaly action is
\begin{equation}\label{GuWenanomaly}
S_5(C)=\frac12\int_P C\cup_1 C = \frac12 \int_P Sq^2 C,
\end{equation}
where $Sq^2: H^3(P,\ZZ_2)\ra H^5(P,\ZZ_2)$ is the Steenrod square. This action is invariant under $C\mapsto C+\delta\beta$, $\beta\in C^2(P,\ZZ_2)$ when $P$ is closed. When $P$ has a nonempty boundary $X$, the action varies as follows:
\begin{equation}\label{5dvariation}
S_5(C+\delta\beta)-S_5(C)=\frac12\int_X \left(C\cup_2 \delta\beta + \beta\cup\beta+\beta\cup_1 \delta\beta\right).
\end{equation}
Note that the variation vanishes when $\delta\beta=0$ and $X$ is a spin 4-manifold. This means that the variation of $S_5$ can be interpreted as an 't Hooft anomaly for a 3+1d bosonic phase which has a global 2-form $\ZZ_2$ symmetry on a spin 4-manifold.

As usual, the anomaly implies that the global 2-form $\ZZ_2$ symmetry acts projectively on the Hilbert space of the bosonic theory associated to a compact 3-manifold $Y$. The 2-cocycle on $Z^2(Y,\ZZ_2)$ corresponding to this projective action is computed in Appendix \ref{descent} and turns out to be
\begin{equation}\label{twococycle3d}
\int_Y \beta\cup_1\beta'.
\end{equation}
This is a symmetric bilinear form on $Z^2(Y,\ZZ_2)$,  and we know from the previous section that its quadratic refinements correspond to spin structures on $Y$. Thus once we fixed a spin structure $\Eta$ on $Y$, we can impose a Gauss law constraint selecting the states in the fermionic Hilbert space for $\Eta$:
$$
U_\beta \vert\Psi,\Eta\rangle=(-1)^{Q_\Eta(\beta)}\vert \Psi,\Eta\rangle
$$
We also identify the fermion parity operator $(-1)^F$ with $(-1)^{\int_Y C}$. 

Note that the 2-cocycle (\ref{twococycle3d}) is not invariant under $\beta\mapsto\beta+\delta\lambda$, and neither is $U_\beta$. So the anomaly is more severe than in the 2d case.

\subsection{Supercohomology phases}

To obtain the supercohomology phases of Gu and Wen, we take the bosonic shadow to be the simplest Crane-Yetter-Kauffman-Walker-Wang model \cite{CY,CKY,WW}:
\begin{equation}\label{WW}
S(a,b)= \frac12 \int_X (a\cup\delta b+b\cup b+b\cup_1\delta b),
\end{equation}
where $a\in C^1(X,\ZZ_2), b\in C^2(X,\ZZ_2)$ are subject to gauge symmetries
\begin{equation}\label{WWgaugesymmetry}
a\mapsto a+\delta f,\quad b\mapsto b+\delta\lambda,\quad f\in C^0(X,\ZZ_2),\ \lambda\in C^1(X,\ZZ_2). 
\end{equation}
The global 2-form $\ZZ_2$ symmetry acts by shifting $b\mapsto b+\beta$,
$\beta\in Z^2(X,\ZZ_2)$. This transformation shifts the action by
\begin{equation}\label{GuWenAnomaly}
\frac12 \int_X \beta\cup\beta=2\pi \frac12 \int_X w_2\cup [\beta],
\end{equation}
where $w_2\in H^2(X,\ZZ_2)$ is the 2nd Stiefel-Whitney class. If $X$ is a closed spin 4-manifold, then $w_2=0$ and the action is invariant for arbitrary $\beta$. Alternatively, if $X$ is not assumed to be spin, the action is invariant only if we impose a constraint $[Sq^2\beta]=0$.

To gauge this 2-form symmetry, we introduce a 3-form gauge field, i.e. a 3-cocycle $C \in Z^3(X,\ZZ_2)$. We modify the action to
$$
S_{gauged}=\frac12 \int_X (a\cup (\delta b+C)+b\cup b+b\cup_1\delta b+C\cup_2\delta b).
$$
The variation of $S_{gauged}$ under a gauge transformation is independent of $a,b$ and given by (\ref{5dvariation}). Thus the theory has the correct 't Hooft anomaly to be a bosonic shadow of a fermionic theory. 

We can promote this theory to a $G$-equivariant model by replacing $C\mapsto C+\rho(A)$, where $\rho \in Z^3(G,\ZZ_2)$. This does not change the anomaly of the 2-form $\ZZ_2$ symmetry, but introduces an anomaly for $G$. To simply notation, let us denote $Sq^2 C=C\cup_1 C$; then 
$$
Sq^2(C+\rho(A)) = Sq^2 C + Sq^2 \rho(A) +\delta( C\cup_2\rho(A))
$$
The last term is exact and thus does not lead to anomaly (it can be absorbed into a contact term $\frac12 \int_X C\cup_2\rho(A)$ in the action). The first term gives the usual anomaly for the 2-form symmetry, while the second term leads to an anomaly for $G$. This anomaly can be canceled if and only if there exists a 4-cochain $\nu$ on $G$ with values in $\RR/\ZZ$ such that
\begin{equation}\label{GuWen3d}
\delta\nu=\frac12 Sq^2\rho=\frac12 \rho\cup_1\rho.
\end{equation}
Then we can cancel the anomaly by adding the term 
$$
\int_X \left(\nu(A)+\frac12 C\cup_2\rho(A)\right)
$$
to the 4d action. The equation (\ref{GuWen3d}) is the Gu-Wen equation for 3d supercohomology phases. 

Before gauging the 2-form $\ZZ_2$ symmetry, the model (\ref{WW}) has loop observables and surface observables. The surface observable localized on a 2d submanifold $\Sigma\subset X$ is $V_\Sigma=\exp(\pi i\int_\Sigma b)$. It is invariant under the gauge symmetry (\ref{WWgaugesymmetry}). It is also charged under the global 2-form $\ZZ_2$ symmetry:
$$
V_\Sigma\mapsto V_\Sigma \exp(\pi i\int_\Sigma\beta),\quad \beta\in Z^2(X,\ZZ_2).
$$
The loop observable localized on a 1d submanifold $\gamma\subset X$ is $W_\gamma=\exp(\pi i \int_\gamma a)$. It is invariant under the gauge symmetry (\ref{WWgaugesymmetry}). When $\gamma=\partial\hat\Sigma$ for some 2-chain $\hat\Sigma$, this loop observable generates the 2-form gauge symmetry with a parameter $\beta_{\hat\Sigma}\in C^2(X,\ZZ_2)$ which is Poincar\'{e} dual to $\hat\Sigma$.   
After gauging the 2-form $\ZZ_2$ symmetry, $V_\Sigma$ is not longer an observable, because it is not gauge-invariant. The loop observable $W_\gamma$ vanishes if $\gamma$ is homologically nontrivial (this follows from the fact that $W_\gamma$ is charged under the global 1-form symmetry $a\mapsto a+\lambda$ for $\lambda\in Z^1(X,\ZZ_2)$), while for homologically trivial $\gamma$ is a generator of a 2-form gauge transformation and therefore is $1$ when inserted into any correlator. The conclusion is that gauging the 2-form $\ZZ_2$ symmetry leads to a theory without any nontrivial observables except the partition function, which depends on the spin structure as well as the $G$ gauge field $A$. This suggests that the gauged theory is a fermionic SPT.\footnote{To establish this, one also needs to prove that the partition function is nonzero on any spin 4-manifold.}

\begin{figure}
\begin{center}
\includegraphics[]{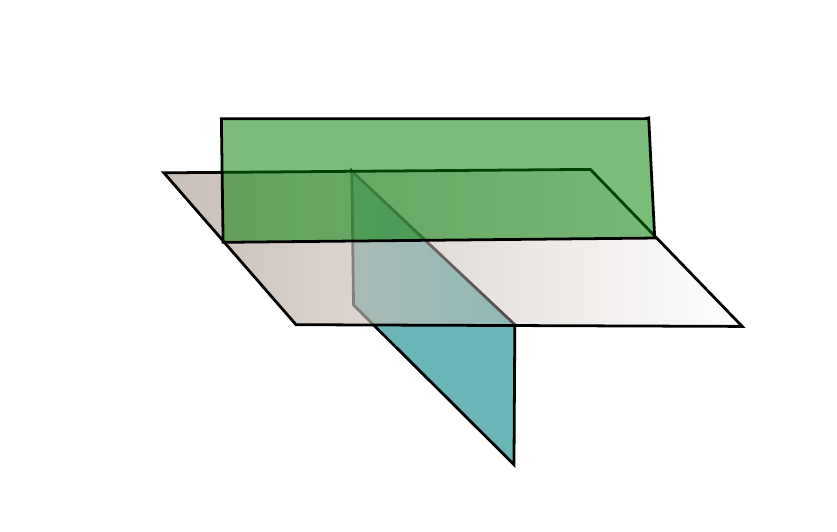}\label{A3}
\caption{A picture of the F-junction or $A_3$ singularity, where four zippers meet. With the x axis along the blue-grey junction and the y axis along the green-grey junction, the planes $x+y = c$ cut through this picture to give a movie of the F move as we vary $c$ through zero.}
\end{center}
\end{figure}

Let us discuss the physical significance of the 3-cochain $\rho \in Z^3(BG,\ZZ_2)$. In a fixed gauge of $A$ at a particular instant of time, we see a network of $G$ domain walls. We will denote by $\dw_g$ the codimension-1 domain wall labeled by $g\in G$.  The SPT ground state should be invariant under the reconnection of the domain wall network. There are several kinds of defects in different dimensions corresponding to different degrees of group cocycles. For example, there is a string-like ``zipper" $\zp_{g,h}$ where the domain walls $\dw_g$, $\dw_h$, and $\dw_{(gh)^{-1}}$  meet. There is also a particle-like fusion junction $\athree_{g,h,k}$ where four zippers meet we call the F-junction or $A_3$ singularity. This is because if we choose a foliation of space by planes transverse to the F-junction, as we scan across we see a movie of the ``F move" or associator where one would apply the F symbol in tensor category theory. These are particle-like objects, and $\rho(g,h,k)$ can be thought of as a way of assigning fermion parity to F-junctions: some are fermionic, some are bosonic.\footnote{We assume here that a zipper does not carry a nontrivial 1d FSRE phase, and thus its endpoints do not have Majorana zero modes. We will discuss zippers with Majorana zero modes later.} We will see this interpretation is natural from the Atiyah-Hirzebruch spectral sequence later. It is exactly analogous to the 2+1D situation where the analogous 2-cocycle $\rho(g,h)$ defines the fermion parity of the triple junction of domain walls. 

The fact that $\rho$ is a 3-cocycle follows from the conservation of fermion number (mod 2), if we assume that the fermion number is the sum of fermion numbers of the F-junctions. Indeed, consider a 4-simplex $T$ whose boundary $\partial T$ consists of five 3-simplices and is homeomorphic to a 3-sphere. The dual of this triangulation of $\partial T$ contains 10 zippers (dual to 2-simplices of $\partial T$) meeting at four F-junctions (dual to 3-simplices of $\partial T$). If the dual of every 1-simplex of $\partial T$ is a domain wall labeled with an element of $G$, then the F-junctions are labeled by three elements of $G$ and have fermion parity determined by $\rho$. On the other hand, since $\partial T$ is a boundary of a 4-simplex, the net fermion number of this configuration of domain walls must vanish mod 2. This is equivalent to the condition $\delta\rho=0$. Alternatively, we can require the fermion parity of a network of domain walls to remain unchanged under 3d Pachner moves. This leads to the same condition on $\rho$. 

Since every zipper has two ends, we can flip the fermion parity of each end without changing the fermion number of the whole network. But this changes the 3-cocycle $\rho$. If the fermion parity of the endpoint of $\zp_{g,h}$ is shifted by $f(g,h)\in\ZZ_2$, it is easy to see that the fermion numbers of the F-junctions change according to
$$
\rho \mapsto \rho + \delta f.
$$
The class $[\rho]\in H^3(BG,\ZZ_2)$ is unchanged.

\subsection{The ground states of the CYKWW model}

The ground-states of the model (\ref{WW}) can be constructed by categorifying the string-net approach \cite{WW}. Roughly speaking, instead of a spherical fusion category, one needs to take a spherical semi-simple monoidal 2-category. Unfortunately, there is no generally accepted definition of this object, and consequently there is no completely general method of constructing 4d TQFTs. But there is a well-understood special case, the CYKWW model \cite{CY,CKY,WW}, and the model (\ref{WW}) belongs to this class. The input of the CYKWW construction is a braided fusion category $\cC$ whose objects represent boundary defect lines for a particular boundary condition. In the present case, the bulk TQFT has a 2-form $\ZZ_2$-symmetry, so we expect that there is an invertible line defect on the boundary which we denote $\psi$ and which satisfies $\psi\circ\psi\simeq 1$. It is a fermion and therefore must have topological spin $-1$. This encodes the fact that the 2-form symmetry has a nontrivial anomaly. If we assume that there are no other irreducible objects in the braided fusion category, then $\cC$ is equivalent to the category of super-vector spaces. 

Let $Y$ be a closed oriented 3-manifold with a triangulation. The CYKWW construction describes the ground states of the model (\ref{WW}) as linear combinations of states $\vert B\rangle$, where $B\in Z^2(X,\ZZ_2)$. In the Poincare-dual picture, $B$ is a network of $\psi$ line defects. A 1-form gauge transformation $B\mapsto B+\delta\lambda$ corresponds to a local rearrangement of the string network. A general state has the form
$$
\vert \Psi\rangle=\sum_B C_B \vert B\rangle.
$$
The string-net Hamiltonian is a commuting projector Hamiltonian whose ground states are distinguished by the way their components transform under a re-arrangement of the string network. Namely, under $B\mapsto B+\delta\lambda$,  $\lambda\in C^1(Y,\ZZ_2)$, one must have
\begin{equation}\label{Ctransf}
C_{B+\delta\lambda}=(-1)^{\int_Y \lambda d\lambda+\delta\lambda\cup_1 B } C_B.
\end{equation}
The explanation for this rule is the following. The category of supervector spaces occurs as a subcategory of the category of bulk line defects for the toric code. Specifically, the line defect $\psi$ can be identified with the $\psi$ of the toric code. Each configuration of $\psi$ lines in $Y$ can be viewed as a network of $\psi$ lines in the toric code, or equivalently as the toric code coupled to a 2-form $\ZZ_2$ gauge field $B$. This gauge field is associated to the anomalous 1-form $\ZZ_2$ symmetry of the toric code whose generator is $\psi$. Rearranging the $\psi$ lines is equivalent to 1-form gauge transformations $B\mapsto B+\delta\lambda$. The rules of the 3d string-net construction tell us that the coefficients $C_B$ transform in the same way as the partition function of the toric code, i.e. (\ref{anomaly3d}). This gives (\ref{Ctransf}).

The transformation rule (\ref{Ctransf}) makes it clear that the number of linearly independent ground states is given by $|H^2(Y,\ZZ_2)|$.
A natural basis in the space of ground states is labeled by spin structures on $Y$. Namely, given a spin structure $\Eta$, we let
$$
\vert\Psi,\Eta\rangle=\sum_B (-1)^{Q_\Eta(B)}\vert B\rangle
$$
After we gauge the 2-form $\ZZ_2$ symmetry, $\vert\Psi,\Eta\rangle$ gives rise to a unique ground state of the 3d FSRE on the spin manifold $(Y,\Eta)$.

\subsection{More general 3d FSRE phases}

The supercohomology phases do not exhaust all possible 3d FSRE phases. There are several ways to see this. For example, one may ask if a zipper $\zp_{g,h}$ (a junction of three domain walls $\dw_g$, $\dw_h,$ and $\dw_{(gh)^{-1}}$) may carry the nontrivial 1d FSRE, i.e. the Kitaev string. The endpoint of such a zipper would have an odd number of Majorana zero modes. Such a phase would be characterized by a new parameter $\sigma(g,h)\in\ZZ_2$ which tells us whether $\zp_{g,h}$ carries the Kitaev string or not. This parameter must be a 2-cocycle. Indeed, consistency requires an even number of Majorana zero modes at each $A_3$ singularity, which is equivalent to the 2-cocycle condition on $\sigma$. 

There is an ambiguity in the definition of $\sigma(g,h)$. The zipper is a place where three domain walls meet. We can attach to the boundary of the domain wall $\dw_g$ a closed Kitaev string; this does not affect any observables, like degeneracies and fermion parities (because the boundary of every domain wall is closed and can be contracted to a point), but it shifts the 2-cocycle $\sigma(g,h)$ by a coboundary.\footnote{One may ask if the boundary of a domain wall can have gapless modes with a nonzero chiral central charge. This would lead to a new parameter $\tau(g)\in\ZZ$ which is easily seen to be a homomorphism from $G$ to $\ZZ$. Since we assumed that $G$ is finite, this parameter vanishes. But this is an interesting possibility if $G$ is infinite and should lead to a new class of fermionic SPT phases.}

When $\sigma$ is nonvanishing, the constraint on the 3-cochain $\rho$ is modified. To see how this comes about, let us try to guess the contibution $F_K$ of Kitaev strings to the fermion number. It is clear that such a contribution can be present whenever $[\sigma(A)]\in H^2(Y,\ZZ_2)$ is nonzero, because this means that there are Kitaev strings wrapping noncontractible loops on $Y$.  In fact, $[\sigma(A)]$ is the Poincare-dual of the homology class of the Kitaev strings. $F_K$ must depend linearly on the spin structure on $Y$. Indeed, shifting the spin structure $\Eta$ by a 1-cocycle $\alpha$ should shift by one the fermion number of a Kitaev string wrapping a curve $\gamma$ if and only if $\int_\gamma\alpha=1$ (since this is when the spin structure induced on $\gamma$ by the ambient spin structure $\Eta$ is flipped by the shift $\Eta\mapsto \Eta+\alpha$)\footnote{See \cite{TF} for a description of how the spin structure (discretized as a Kastelyn orientation + dimer covering) is implemented microscopically on the Kitaev string.}. Hence we expect 
$$
F_K(\Eta+\alpha)=F_K(\Eta)+\int_Y \alpha\cup \sigma(A).
$$
The quadratic function $Q_\Eta(B)$ depends on $\Eta$ as expected, provided we identify $B=\sigma(A)$. We therefore propose that
$$
F_K(\Eta)=Q_\Eta(\sigma(A)).
$$

An important property of $Q_\Eta(\sigma(A))$ is that it is not invariant under replacing the 2-cocycle $\sigma(A)$ with a cohomologous one. In the language of Kitaev strings, this means that $F_K$ changes when Kitaev strings are reconnected. Since the net fermion number must be conserved, we propose that whenever the 2-cocycle $\sigma$ is shifted by $\delta\lambda$, the 3-cocycle $\rho$ determining the $\athree$ contribution to the fermion number $F$ shifts:
\begin{equation}\label{rhoshift}
\rho\mapsto\rho+\lambda\cup\delta\lambda+\delta\lambda\cup_1 \sigma. 
\end{equation}
Then this compensates for the transformation rules of $Q_\zeta$ so
$$
F=Q_\Eta(\sigma(A))+\int_Y \rho(A)
$$
is gauge-invariant.

The constraint $\delta\rho=0$ is not invariant under the shift (\ref{rhoshift}), which is not surprising, since the fermion number is now carried both by the junctions $\athree_{g,h,k}$ and the zippers $\zp_{g,h}$. But a modified constraint
\begin{equation}\label{rhosigmaeq}
\delta\rho=\sigma\cup\sigma
\end{equation}
is invariant under the shift (\ref{rhoshift}) accompanied by  $\sigma\mapsto\sigma+\delta\lambda$. 

We can reach the same conclusion assuming the relation between 3d FSRE phases and spin-cobordism. The Atiyah-Hirzebruch spectral sequence converging to the spin-cobordism of $BG$ indicates that there should be three parameters: $\nu\in C^4(BG,\RR/\ZZ)$, $\rho\in C^3(BG,\ZZ_2)$ and $\sigma\in C^2(BG,\ZZ_2)$. These parameters must satisfy constraints which at linearized level are simply $\delta\nu=\delta\rho=\delta\sigma=0$, but have corrections which are encoded in the differentials of the spectral sequence.
The spectral sequence immediately implies that the constraint on $\sigma$ is not modified by the differentials, i.e. $\delta\sigma=0$, in agreement with the physical argument above, but that other constraints are modified. The 1st differential in the spectral sequence is known to be the Steenrod square $Sq^2$, suggesting that the equation for $\rho$ is modified to (\ref{rhosigmaeq}).

The equation $\delta\nu=0$ is modified at leading order as well, to the Gu-Wen equation $\delta\nu=\frac12 Sq^2\rho=\frac12 \rho\cup_1\rho,$, but it must receive higher-order modifications as well, in order to be consistent with (\ref{rhosigmaeq}). It is shown in Appendix \ref{Eanomalies} that there is an essentially unique modification  of the Gu-Wen equation consistent with (\ref{rhosigmaeq}). We thus propose that 3d FSRE phases are classified by solutions of the equations
 \begin{equation}\label{genGuWen3d}
 \delta\nu=\tilde Sq^2_\pm(\rho,\sigma) =\frac12 \rho\cup_1\rho+\ldots,\quad \delta\rho=\sigma\cup\sigma,\quad \delta\sigma=0,
 \end{equation}
where dots denote terms which depend only on $\sigma$ which are required to make the r.h.s. of the first equation closed mod integers. We give an explicit formula for $\tilde Sq^2_\pm$ in the appendix.
There are also several nontrivial identifications on the set of solutions. The abelian group structure is also highly nontrivial, since the equations appear nonlinear \cite{MorganBrumfiel2}. Suffice it to say that the space of solutions has an obvious subgroup corresponding to solutions of the form $(\nu,0,0)$, where $\nu\in Z^4(BG,\RR/\ZZ)$. This subgroup consists of bosonic SRE phases. Taking a quotient by this subgroup leads to a more manageable object (the group of 3d FSRE phases modulo bosonic SRE phases) which consists of equivalence classes of pairs $(\rho,\sigma)$ satisfying $\delta\rho=\sigma\cup\sigma,$ $\delta\sigma=0$. The equivalence relation arises from a gauge symmetry
$$
\sigma\mapsto \sigma+\delta\lambda,\quad \rho\mapsto\rho+\delta\beta+\lambda\cup\delta\lambda+\delta\lambda\cup_1\sigma,\quad \lambda\in C^1(BG,\ZZ_2),\ \beta\in C^2(BG,\ZZ_2).
$$
The abelian group structure on these equivalence classes is easily guessed:
\begin{equation}\label{grouplaw3dFSRE}
 (\rho,\sigma)+(\rho',\sigma')=(\rho+\rho'+\sigma\cup_1\sigma',\sigma+\sigma').
 \end{equation}
Indeed, since
$$
(\sigma+\sigma')\cup(\sigma+\sigma')=\sigma\cup+\sigma'\cup\sigma'+\delta(\sigma\cup_1\sigma'),
$$
the r.h.s. of eq. (\ref{grouplaw3dFSRE}) satisfies the equation (\ref{rhosigmaeq}) provided the l.h.s. does. It is straightforward to verify the group axioms. 
 
\subsection{Bosonization of 3d FSRE phases and 3-group symmetry}

Let us interpret the above proposal for the classification of 3d FSRE phases in terms of their bosonic shadows. It was argued in \cite{GK,BGK} that one can construct a 3d fermionic system from a 3d bosonic system with a global $\ZZ_2$ 2-form symmetry provided this 2-form symmetry has a suitable 't Hooft anomaly. A natural generalization of this construction is to combine it with gauging some other symmetries of the bosonic system. Now, suppose the symmetry of the bosonic system is not simply a product of the 2-form $\ZZ_2$ symmetry and other symmetries, but a more general structure. Specifically, since the general 3d  FSRE phases are supposed to contain both a condensate of fermionic particles and a condensate of Kitaev strings, we are led to consider bosonic shadows with both a 2-form $\ZZ_2$ symmetry and a 1-form $\ZZ_2$ symmetry.  Particles will be associated with generators of the 2-form symmetry, while strings will be associated with generators of the 1-form symmetry. 

In general, when a field theory has 0-form, 1-form and 2-form symmetries, the whole symmetry structure is described by a 3-group. A general 3-group is quite a complicated object, but it simplifies when we ignore 0-form symmetries. In that case, the 3-group is characterized by its 1-form symmetry group $G_1$, its 2-form symmetry group $G_2$, and a Postnikov class taking values in $H^4(BG_1,G_2)$. In the present case $G_1=G_2=\ZZ_2$, and $H^4(B\ZZ_2,\ZZ_2)=\ZZ_2$, so there is only one nontrivial possibility for the Postnikov class. If the Postnikov class vanishes, the 3-group is simply a product of 1-form and 2-form symmetries. If it is nontrivial, the 2-form gauge field $B$ is still closed, while the 3-form gauge field $C$ satisfies
\begin{equation}\label{threegroup}
\delta C=B\cup B.
\end{equation}
Note the similarity with Eq. (\ref{rhosigmaeq}). 

The modified Bianchi identity (\ref{threegroup}) gives rise to a modified group law for global symmetry transformations. To derive the group law, we assume that 2-form symmetry transformations leave $B$ invariant and shift $C$:
$$
B\mapsto B,\quad C\mapsto C+\delta\beta,\quad \beta\in C^2(X,\ZZ_2),
$$
while 1-form symmetry transformations shift $B$:
$$
B\mapsto B+\delta\lambda,\quad\lambda\in C^1(X,\ZZ_2).
$$
Then (\ref{threegroup}) requires $C$ to transform as follows under 1-form gauge symmetry:
$$
C\mapsto C+\lambda\cup\delta\lambda+\delta\lambda\cup_1 B.
$$
Now consider the effect of two 1-form symmetry transformations with parameters $\lambda$ and $\lambda'$ on the configuration $B=0,C=0$. We get
$$
B=\delta(\lambda+\lambda'),\quad C=\lambda' \delta\lambda'+\lambda\delta\lambda+\delta\lambda'\cup_1\delta\lambda.
$$
The first equation shows that this is equivalent to a 3-group symmetry transformation with a 1-form symmetry transformation $\lambda+\lambda'$ and an undetermined 2-form symmetry transformation with a parameter $\beta(\lambda,\lambda')$. The second equation then implies that 
$$
\beta(\lambda,\lambda')=\lambda\cup\lambda'. 
$$
Specializing to closed $\beta$ and $\lambda$, we conclude that the group law for global 3-group symmetry transformations is
\begin{equation}\label{Egrouplaw}
(\beta,\lambda)+(\beta',\lambda')=(\beta+\beta'+\lambda\cup\lambda',\lambda+\lambda'),\quad \beta\in Z^2(X,\ZZ_2),\ \lambda\in Z^1(X,\ZZ_2).
\end{equation}

Consider now coupling the bosonic theory to a $G$ gauge field $A$ by letting $C=\rho(A)$ and $B=\sigma(A)$ for some $\rho\in C^3(BG,\ZZ_2)$ and $\sigma\in C^2(BG,\ZZ_2)$. Since $B$ must be closed, $\sigma$ must be a 2-cocycle. Since $C$ satisfies (\ref{threegroup}), we must subject $\rho$ to (\ref{rhosigmaeq}).
To get the 3d FSRE, we gauge the 3-group symmetry, while keeping $A$ fixed. To ensure gauge-invariance with respect to $G$ gauge transformations, we need to impose further constraints on the data $\rho$ and $\sigma$, like the first equation in (\ref{genGuWen3d}). 

Thus our proposal for 3d bosonization can be formulated as follows: every fermionic theory has a bosonic shadow with a global 3-group symmetry as above (we will denote this 3-group $E$) and an 't Hooft anomaly $\tilde Sq_\pm(C,B)$ (see Appendix F for the definition of the latter). In particular, we propose that every 3d FSRE can be constructed in this way. This construction is more general than that proposed in \cite{GK}. To see this, note that the 2-form $\ZZ_2$ symmetry is a proper subgroup of the 3-group symmetry, so we are free to gauge it first and get a 3d fermionic phase as in \cite{GK}. But this fermionic phase is not an FSRE yet: it has nontrivial observables charged under the global 1-form $\ZZ_2$ symmetry (this symmetry is what remains of the 3-group symmetry after we gauge the 2-form symmetry). To get an FSRE we must also gauge this 1-form symmetry. The order of the steps in this two-step procedure cannot be reversed, since the 1-form symmetry is not a subgroup of the 3-group symmetry, and cannot be gauged without gauging the whole 3-group. 

The description of the 3-group gauging as a two-step process makes it intuitively clear that the resulting phase is a fermionic phase, since the spin structure is introduced already at the first step. But it is not clear  that a further geometric structure is not needed at the second step. The question boils down to computing possible anomalies for a 1-form $\ZZ_2$ symmetry in a fermionic theory, taking into account that the 2-form gauge field $B$ satisfies the constraint $Sq^2[B]=0$. It is shown in the Appendix that no anomaly is possible, and thus the 3-group symmetry with the above anomaly can always be gauged on a spin 4-manifold.

\section{Bosonic shadows of 3d FSRE phases}

\subsection{2-Ising Theory}

The goal of this section is to construct a 3+1d TQFT which is a 3+1d analogue of the Ising TQFT in 2+1d, and has global 3-group symmetry $E$. Physically we imagine a gapped superconductor with fermionic charges and vortex lines which terminate at Majorana zero modes on the boundary. Since the Ising category describes the behavior of Majorana zero modes, it will also describe the behavior of these charges and vortex strings.

This TQFT is nonabelian, so we will need an algebraic approach to construct it. This approach is a 4d analog of the Turaev-Viro construction and takes a monoidal 2-category as an input \cite{4dTQFT}. From the physical viewpoint, this monoidal 2-category describes boundary defects for a particular topological boundary condition. 

Since the 4d TQFT has both 2-form and 1-form $\ZZ_2$ symmetries, it contains a codimension-3 defect (the generator of the 2-form $\ZZ_2$ symmetry) and a codimension-2 defect (the generator of the 1-form $\ZZ_2$ symmetry). Let us denote by $\psi$ the fusion of the codimension-3 defect with the boundary. It is a line defect on the boundary, or equivalently a line defect on the ``transparent'' surface defect. Algebraically, the `transparent'' defect is the identity object $E$ of the monoidal 2-category, and thus $\psi\in \Hom(E,E)$. The fusion of the codimension-2 defect with the boundary gives us another object which we denote $O$. We have the fusion algebra
$$
O\otimes O\simeq E.
$$
We postulate that there are no further indecomposable defects in the 4d TQFT, and thus every object in the monoidal 2-category is a direct sum of several copies of $E$ and $O$ (this is a natural assumption since gauging both 2-form and 1-form symmetries should lead to a theory with no nontrivial observables). 

Next we need to describe morphism categories. $\Hom(E,E)$ is a braided fusion category, and by assumption it is generated by $\psi$ and the identity object $1$. The $\ZZ_2$ fusion rule $\psi\circ\psi\simeq 1$ means that $\Hom(E,E)$ is equivalent to the category of $\ZZ_2$-graded vector spaces as a fusion category. There are two braided structures on it: one corresponds to the usual tensor product, and the other one to the supertensor product. They correspond to two possible anomalies for the 3-form symmetry: the trivial one and the one with the anomaly action $\int Sq^2C$. We need the latter option, so that $\Hom(E,E)$ is equivalent to the category of supervector space as a braided fusion category. Assuming that $O$ is its own dual, we can also compute $\Hom(O,O)$:
$$
\Hom(O,O)\simeq \Hom(O\otimes O,E)\simeq \Hom(E,E)= \langle 1, \psi\rangle.
$$

Finally, we need to describe the categories $\Hom(E,O)$ and $\Hom(O,E)$. We postulate that both $\Hom(O,E)$ and $\Hom(E,O)$ are non-empty and each of them has a single irreducible object which we denote $\sigma$. This means that the surface defect $O$ can terminate on the boundary. Nevertheless, $O$ is not equivalent to $E$, because $\sigma$ is not invertible. We postulate the simplest non-invertible fusion rule:
\begin{equation}\label{sigmafusion}
\sigma \circ \sigma = 1 \oplus \psi.
\end{equation}
We also necessarily have 
$$
\sigma\circ\psi\simeq\sigma,\quad \psi\circ\sigma\simeq\sigma,
$$
because $\psi$ is invertible.


To complete the construction of the monoidal 2-category we need to specify all the associator morphisms and the pentagonator 2-morphisms \cite{4dTQFT}. This is facilitated by the fact that the monoidal 2-category we are constructing has a very special form: its data are equivalent to those of a braided $\ZZ_2$-crossed braided category \cite{Turaev}. This is a $\ZZ_2$-graded fusion category $\cC=\cC_0+\cC_1$ with a compatible $\ZZ_2$-action and additional data which generalizes braiding and reduces to it when the $\ZZ_2$-action is trivial. In our case, 
$\cC_0=\Hom(E,E)$, and $\cC_1=\Hom(E,O)$, and the $\ZZ_2$ action is trivial.  Thus $\cC$ is an Ising braided fusion category (and therefore is a braided  $\ZZ_2$-crossed category). All possible braided fusion structures on an Ising category are known, and it turns out there are eight inequivalent ones, naturally labeled by a complex number $\kappa$ such that $\kappa^8=-1$. 

To any braided $\ZZ_2$-crossed category one can associate a 4d TQFT using a generalization of the CYKWW construction \cite{CuiThesis}. We will call the 4d TQFT obtained by taking the braided Ising category as an input for this construction a 2-Ising model. We will show that the 3-group symmetry generated by $O$ and $\psi$ is isomorphic to $E$ and has the correct anomaly to be a bosonic shadow. We propose that the bosonic shadows of all 3d FSREs can be obtained by taking the 2-Ising model and coupling it to a background $G$ gauge field while keeping the anomalies intact. Below we provide some evidence for this.

\subsubsection{State Sum}

In this section we describe the state sum for the 2-Ising model. The state sum is a sum over colorings of a triangulation of $X$ with a fixed branching structure. A coloring is an assignment of (simple) objects to edges, (simple) morphisms to triangles, and (simple) 2-morhisms to tetrahedra. The weight of each coloring is a product over the 15j symbol in each 4-simplex. The partition function on $X$ is a sum of weights over all colorings. Let us spell out what this means for 2-Ising.

\begin{itemize}

\item Edges are labeled either $X_{01} = E$ or $X_{01} = O$.

\item At a triangle, we assign a morphism in the fusion space (a category) $\Hom(X_{01} \otimes X_{12}, X_{02})$ where we have labeled the triangle according to the chosen branching structure. In particular, we have either two $O$'s, in which case the only morphism in the fusion space is $\sigma$; or no $O$'s, in which case the morphism may be either 1 or $\psi$.

\item At a tetrahedron, it is useful to imagine the dual picture, shown in Fig 1, where six sheets are meeting, with a 1, $\psi$, or $\sigma$ on each of four fusion junctions which meet at a point in the center. At this point, we need to have something gluing together the fusion junctions. The rules for this is precisely the same as in the usual Ising category. That is, we can forget the sheets and just think of this as a junction of 1, $\psi$, and $\sigma$ lines. Choosing a resolution of the 4-valent vertex into two 3-valent vertices defines a basis for this fusion space.

\item The coloring around a 4-simplex is a collection of $O$-sheets and $\psi$ and $\sigma$ lines around its boundary, a 3-sphere. The branching structure defines a framing of this 3-sphere and we can use the rules of the Ising category \cite{KitaevAnyons} to evaluate it to a number. This defines the ``15j'' symbol of \cite{CY} and is the weight of the coloring in the state sum. See Fig 2.

\begin{figure}\label{4simplex}
\begin{center}
\includegraphics[width=8cm]{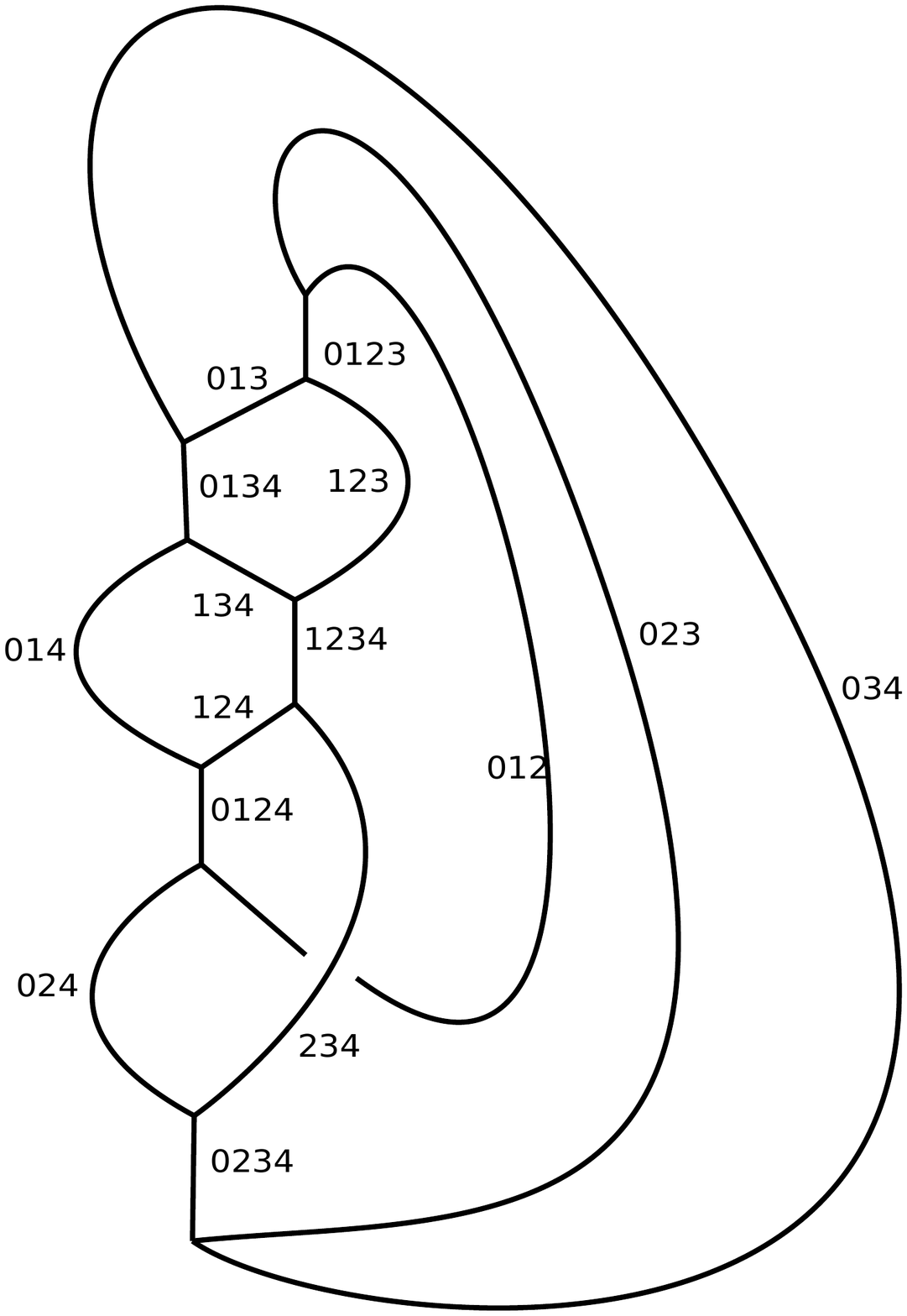}\label{flat4simplex}
\caption{This image, essentially a reproduction of Fig 16 from \cite{CuiThesis}, represents the boundary of the 4-simplex, considered as a triangulation of the 3-sphere. We have flattened the image onto the page using a framing induced from the branching structure, which gives us a labeling of vertices 0 through 4. This picture is actually Poincar\'e dual in the 3-sphere to that 4-simplex, with (most) edges here representing triangles of the vertex-ordered 4-simplex. This is so the graph depicts the labeling of triangles in $X$ by line objects. Representing triangles, these edges are  labeled by triples of vertices, and there are 5 choose 3 of those. There are some extra edges where we have resolved 4-way intersections to make the graph trivalent. These are labeled by tetrads of vertices and coincide with the 5 choose 4 tetrahedra of the dual 4-simplex. The state sum gives us a labeling of these edges by 1, $\psi$, and $\sigma$, and the rules of the Ising braided fusion category of \cite{KitaevAnyons} gives us a way to evaluate this picture to a number. This defines the 15j symbol.}
\end{center}
\end{figure}


\end{itemize}

\subsubsection{$E$ Symmetry}

It is useful to encode the state sum as a sum over cochains. We define the following cochains in spacetime $X$:
\begin{itemize}
\item $\epsilon_1 \in C^1(X,\ZZ_2)$ is Poincar\'e dual to the $O$ worldvolume.
\item It follows from the fusion rules that $d\epsilon_1$ is Poincar\'e dual to the $\sigma$ worldsheet.
\item $\epsilon_2 \in C^2(X,\ZZ_2)$ is Poincar\'e dual to the $\psi$ worldsheet.
\item It follows from the rules of the Ising category that
\begin{equation}\label{hopfconstraint}
d\epsilon_2 = \epsilon_1 \cup d\epsilon_1.
\end{equation}
\end{itemize}
This last point deserves some elaboration. We can imagine each configuration in the state sum on $X$ as a movie of fluctuating $\sigma$ and $\psi$ lines and $O$ surfaces which evolve according to the local moves of the usual Ising category, except for the $O$ surfaces making the $\sigma$ worldsheet always a boundary and inducing the into-the-page framings for the evaluation of the Ising $R$ and $F$ matrices.

On the boundary of a 4-ball in $X$ we see a snapshot of the action. In this snapshot, we may have two $\sigma$ lines in a Hopf-link formation with $O$ surfaces defining the into-the-page framing as shown in Fig. 3. According to the rules of the Ising category (see e.g.~\cite{KitaevAnyons}), this configuration can only be filled into the 4-ball if those $\sigma$ lines have a $\psi$ connecting them.

For this rule to be insured by the local dynamics of the 2-Ising Hamiltonian, the term which creates small discs of $O$ surface must create $\psi$ lines along the intersections of $O$ surfaces. There are also terms which create small loops of $\psi$ line, but these cannot move the endpoints of the $\psi$ lines, which will be where the $O$ surface intersects the $\sigma$ line. From this follows the equation (\ref{hopfconstraint}).

\begin{figure}\label{hopf}
\begin{center}
\includegraphics[width=8cm]{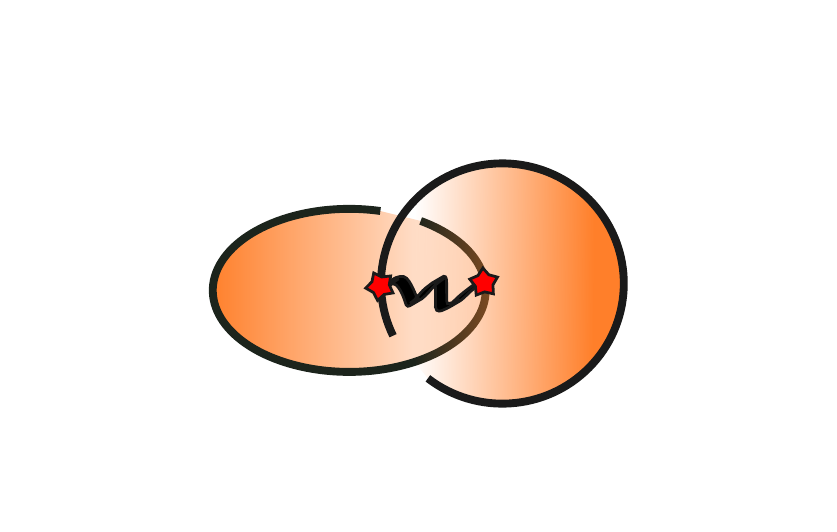}
\caption{A configuration of two $\sigma$ anyons (black circles) in a Hopf link formation. The $\sigma$'s are the boundary of the $O$ surface (orange discs). Where the $\sigma$'s intersect the orange disc (red stars), we have a $\psi$ anyon (wavy black curve) being born.}
\end{center}
\end{figure}

The equation (\ref{hopfconstraint}) implies that the 2-Ising model carries an action of the symmetry 3-group $E$. To see this, note that the global 2-form symmetry acts by  $\eps_2\mapsto\eps_2+\beta$, $\beta\in Z^2(Y,\ZZ_2)$ while leaving $\eps_1$ unchanged, where $Y$ is a spatial 3-manifold. The 1-form $\ZZ_2$ symmetry shifts $\eps_1\mapsto \eps_1+\alpha$, $\alpha\in Z^1(Y,\ZZ_2)$, and to be consistent with (\ref{hopfconstraint}) one must also transform $\eps_2$: $\eps_2\mapsto\eps_2+\alpha\cup\eps_1$. A general symmetry transformation is parameterized by a pair $(\alpha,\beta)\in Z^1(Y,\ZZ_2)\times Z^2(X,\ZZ_2)$ and acts as follows:
\begin{equation}\label{epsilonaction}
\eps_1\mapsto\eps_1+\alpha,\quad \eps_2\mapsto\eps_2+\alpha\cup\eps_1+\beta.
\end{equation}
Performing two consecutive transformation we get the group law Eq. (\ref{Egrouplaw}):
\[(\beta_1,\alpha_1)+(\beta_2,\alpha_2) = (\beta_1+\beta_2+\alpha_1\cup\alpha_2,\alpha_1+\alpha_2).\]
Thus the 2-Ising model is acted upon by the 3-group $E$. As explained in section 4.5, with the proper anomaly such a symmetry is ``fermionic'' in that we can gauge it by introducing a spin structure.

\subsection{Fermion Number}

The relation (\ref{hopfconstraint}) has other interesting consequences. To evaluate the right hand side on a tetrahedron, we need to order the vertices 0, 1, 2, 3 and compute
\[(\epsilon_1 \cup \delta\epsilon_1)(0123) = \epsilon_1(01)(\epsilon_1(12)+\epsilon_1(23)-\epsilon(13)).\]
This quantity depends on the choice of ordering, which we take to be defined by a branching structure on $X$. One way to understand this is that
\[\int \epsilon_1 \cup \delta\epsilon_1 \mod 2\]
computes the mod 2 self-linking number of the $\sigma$ curves with respect to the framing defined by the $O$ surfaces. That is, it equals the mod 2 linking number of the $\sigma$ curve and the curve obtained from $\sigma$ by displacing it a small distance into the piece of $O$ surface that bounds it. The integral of $\epsilon_1 \delta\epsilon_1$ is counting crossings between these two curves, and of course where the crossings are depends on the local framing of space.

The $\psi$ lines are line defects in the 4d theory whose endpoints represent the fundamental fermion. The equation (\ref{hopfconstraint}) then says that the total fermion number of the state is the self-linking of the $\sigma$ loops  framed by $O$. The configurations which appear in the 2-Ising state sum all have even net fermion number, but the number of points where the $\psi$ lines are attached depends crucially on the framing. For example, we can create a Hopf link of $\sigma$ loops without any $\psi$ lines by having an $O$ surface which is a twice-twisted ribbon. See Fig. 4.
\footnote{This diagram is not subject to the rules of the Ising category $S$-matrix because the framing of $\sigma$ induced by the $O$ surface does not extend to any framing of $S^3$. All of the Ising category numbers are computed in 2-Ising by choosing the $O$ surfaces so that they induce into-the-page framings and then computing the wavefunction overlap with the empty picture.} This is a new ingredient for topological order in 3+1D. Quasiparticles can only be the boundary of string operators in one way, but 2-Ising illustrates how the statistics of a quasistring depends on how it is framed by its bounding surface operator.

\begin{figure}
\label{hopf2}
\begin{center}
\includegraphics[width=8cm]{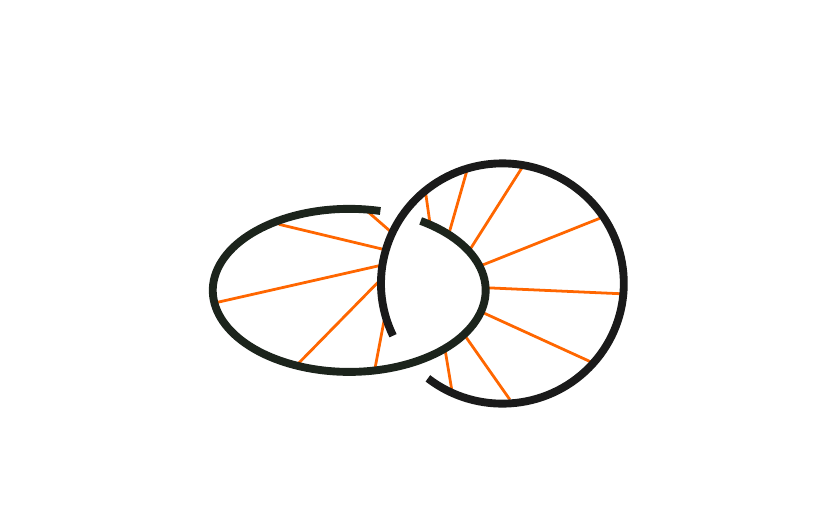}
\caption{A configuration of two $\sigma$ anyons (black circles) in a Hopf link formation given by the boundary of a twice-twisted ribbon of $O$ surface (orange skeleton). With this configuration of the $O$ surface, the self-linking of each component is even, so there is no need for a $\psi$ line connecting them. This contrasts with the into-the-page framed Hopf link we drew above, where the framing induced by the $O$ surface has odd self-linking in each component, so the two components are fermionic and there must be a $\psi$ line connecting them.}
\end{center}
\end{figure}

When we gauge the $E$ symmetry, the $O$ surfaces will proliferate and the $\sigma$ loops will lose their framings. Their density will be measured by a $\ZZ_2$ 2-form $B$ which is Poincar\'e dual to the worldsheet (its integral over a surface counts the number of $\sigma$ strings piercing it). In order for the fermion number $F_K(B)$ of the loops to be well-defined, one needs some geometric input that stands in for the framing of the $\sigma$ loops. From what we have discussed so far, the function has to satisfy
\[F_K(\delta\epsilon_1) = \int \epsilon_1 \cup \delta\epsilon_1 \mod 2,\]
which is a special case of Eq. (\ref{rhoshift}) replacing $\lambda$ by $\epsilon_1$. It is possible to achieve this by framing all of space since this frames all curves so that their mod 2 self-linking numbers are well-defined, but this is not very physical and too restrictive for our goals. As discussed in section 4.4, we can define such an $F_K$ given a spin structure on spacetime. This is very physical, since we wish to describe fermionic systems by gauging the $E$ symmetry. We conclude that the fermion number of the $\sigma$ loops will depends on the spin structure $\Eta$. Let us write the string fermion number $F_{K,\Eta}$.

We can guess the dependence of $F_{K,\Eta}$ on the spin structure $\Eta$ by thinking about the spin structure induced by a framing. As discussed in \cite{kirbytaylor}, one can think of a spin structure in 3d as a mod 2 invariant of any framed curve which increments by 1 when the framing of the curve is twisted. It also flips by $\int \alpha \mod 2$ when the spin structure $\Eta$ is shifted by a $\ZZ_2$ 1-cocycle $\alpha$ to $\Eta + \alpha$. Thus, changing the spin structure is equivalent to twisting certain framings. This also changes the mod 2 self-linking by the same amount, so we find that the fermion number is linear in the spin structure:
\[F_{K,\Eta+ \alpha}(B) = F_{K,\Eta}(B) + \int \alpha\cup B \mod 2.\]
As discussed in Appendix \ref{QEta}, this requirement essentially fixes $F_{K,\Eta}(B)$ to be the function $Q_\Eta(B)$. 

Gauging the $E$ symmetry also frees the fundamental fermion, the endpoint of the $\psi$ lines, turning the 3-coboundary $\delta\epsilon_2$ into a 3-cochain $C$. It follows from Eq. (\ref{hopfconstraint}) that
\[\delta C = B\cup B.\]
This reflects the non-trivial Postnikov class of the 3-group $E$.

\subsection{$G$-crossed 2-Ising}

Now we want to enlarge 2-Ising to a theory with a global $G$ symmetry. We do this by extending $G$, considered as a monoidal 2-category with trivial morphisms, by 2-Ising. The data for this will consist of a group 2-cochain $\sigma \in C^2(G, \ZZ_2)$, a group 3-cochain $\rho \in C^3(G,\ZZ_2)$, and a group 4-cochain $\nu \in C^4(G,U(1))$, satisfying some conditions we presently derive.

Physically, the presence of a global symmetry $G$ means that for every $g\in G$ there is a codimension-1 invertible defect. Their fusion obeys the group law of $G$. If the symmetry is unbroken on the boundary, each such defect gives rise to an invertible surface defect on the boundary which we denote $E_g$. Fusing each of them with $O$ (the generator of the 1-form symmetry), we get another surface defect $O_g$. Obviously, $O_g\simeq O_1\otimes E_g$. Although the fusion of bulk domain walls obeys the group law of $G$, the fusion of $E_g$ and $O_g$ is governed by the group law of an extension of $G$ by $\ZZ_2$. This happens because the termination of a bulk defect is not canonically defined, so for a given $g$ one can always swap $E_g$ and $O_g$. If $\sigma(g_1,g_2)\in \ZZ_2$ is a 2-cocycle describing this extension, then the fusion rule is
\begin{equation}
E_{g_1}\otimes E_{g_2}\simeq E_{g_1 g_2} \otimes O^{\sigma(g_1,g_2)}.
\end{equation}
Associativity of the fusion algebra is equivalent to 
\begin{equation}\label{nu2closed}
\delta \sigma = 0.
\end{equation}

Geometrically, this means the following: a zipper $\zp(g,h)$ is part of the boundary of $O$ surfaces if and only if $\sigma(g,h)=1$. 
Thus, in the gauge where $\sigma$-lines are absent, we must have
\begin{equation}\label{gsigmaconstraint}
\delta\epsilon_1 = \sigma(A),
\end{equation}
where $A \in Z^1(X,G)$ represents the configurations of $G$ labels on objects on edges in the state sum. In the bulk, where an $O$ surface cannot terminate, we postulate that (\ref{gsigmaconstraint}) holds without any restrictions. 

Next we interpret the 3-cochain $\rho\in C^3(G,\ZZ_2)$. Whenever this 3-cochain is nonzero, the $A_3$ singularities where four zippers meet are sources of $\psi$ lines. Thus the 2-cochain $\eps_2$ representing $\psi$-lines must satisfy
\begin{equation}\label{rhoconstraint}
\delta\eps_2=\rho(A)+\eps_1\cup\delta\eps_1.
\end{equation}
The second term is required to ensure that the constraint is invariant under the action of the $E$ symmetry (\ref{epsilonaction}). 

Now let us integrate (\ref{rhoconstraint}) over $Y$, assuming that $Y=\partial X$. Taking into account that all cochains in (\ref{rhoconstraint}) are restrictions of cochains on $X$ and using the Stockes theorem and (\ref{gsigmaconstraint}), we get 
\begin{equation}\label{Xpostnikov}
\int_X\delta\rho(A)=\int_X\sigma(A) \cup \sigma(A).
\end{equation}
Since $A$ and $X$ are arbitrary, we must have
\begin{equation}\label{nupostnikov}
\delta\rho=\sigma\cup\sigma. 
\end{equation}

With the constraints in Eqs. \ref{nu2closed} and \ref{nupostnikov}, the pair $(\sigma,\rho)$ describes precisely a map $BG\ra BE$, where $BE$ is the classifying space for the 3-group $E$. Thus a $G$-crossed 2-Ising model is an equivariantization of the 2-Ising model. 

One more constraint on $\rho$ and $\sigma$ should follow from the topological invariance of the $G$-crossed 2-Ising model. In principle, it can be obtained by evaluating the partition function on a boundary of a 5-ball and requiring it to be $1$ for an arbitrary gauge field $A$ on the 5-ball and arbitrary $\eps_1$ and $\eps_2$ satisfying all the constraints. For a trivial gauge field $A$, this is ensured by the properties of the 2-Ising 15j symbol. Instead of performing this rather formidable computation for general $A$, we can take a short-cut and require the symmetry $G$ to be non-anomalous. Since we embedded $G$ into $E$ by letting $C=\rho(A)$ and $B=\sigma(A)$, this means that the cocycle
\begin{equation}\label{2IsingGuWen}
\tilde Sq^2_\pm(\rho,\sigma)\in C^5(G,\RR/\ZZ),
\end{equation}
must be exact. Here the choice of the sign in $\tilde Sq^2_\pm$ depends on the braiding structure of the 2-Ising theory we use. Thus there must exist a 4-cochain $\nu\in C^4(G,\RR/\ZZ)$ such that
\begin{equation}
\delta\nu=\tilde Sq^2_\pm(\rho,\sigma).
\end{equation}
We propose that the general 3d FSRE with symmetry $G$ can be obtained by gauging the 3-group symmetry $E$ of the $G$-crossed 2-Ising model.


\subsection{Super-Cohomology Phases from $G$-crossed 2-Ising}

When $\sigma=0$, the sector of the $G$-crossed 2-Ising containing $O$ surfaces and $\sigma$-lines decouples, and we can restrict our attention to the networks with $\eps_1=0$. The remaining constraints simplify to
\begin{equation}\label{eps2rhoconstraint}
\delta\eps_2=\rho(A). 
\end{equation}
Thus for a fixed network of $G$ domain walls, we sum over all networks of $\psi$ lines satisfying the following condition: each $A_3$ singularity  $\athree_{g,h,k}$ with $\rho(g,h,k)=1$ is a source for a $\psi$-line, and $\psi$-lines cannot end anywhere else. 

\begin{figure}
\begin{center}
\includegraphics[width=8cm]{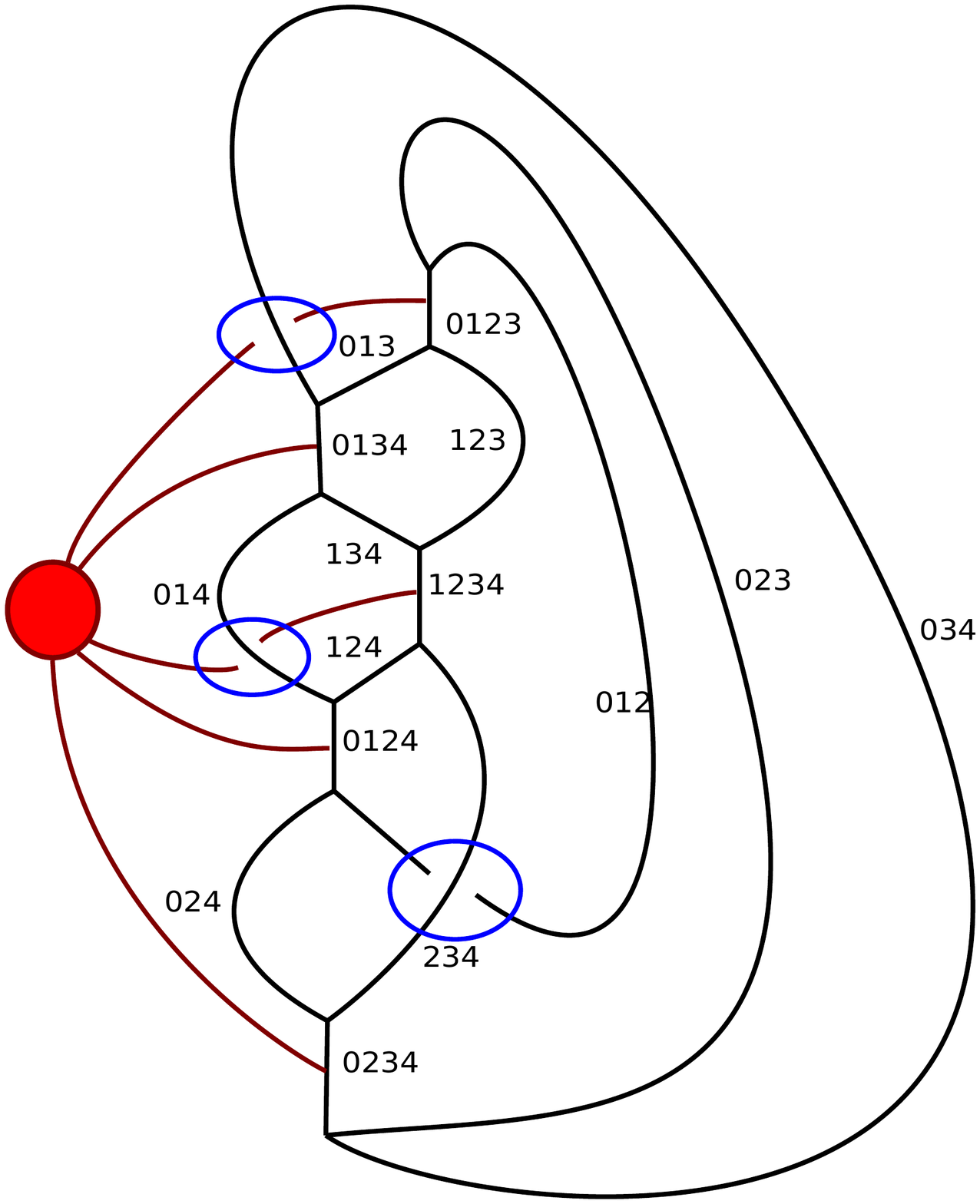}\label{flat4simplex}
\caption{We revisit the 15j symbol in the presence of $C$ with non-zero $\rho$ but $\sigma = 0$. The tetrahedra where $\rho \neq 0$ have a non-conservation of $\psi$ lines, indicated by red curves coming out of the resolved 4-way junctions ($A_3$ singularities) dual to the tetrahedra. The $\psi$ lines go and join ``the condensate'', represented by a red ball which may absorb any number of $\psi$ lines. In evaluating the diagram according to the rules of the Ising category, we get contributions from crossings. The red with black give a sign contribution of $-1$ to power $\epsilon_2(034)\rho(0123)+\epsilon_2(014)\rho(1234) = (\epsilon_2 \cup_1 \rho)(01234)$. The black with black crossing gives a contribution of $-1$ to power $\epsilon_2(012)\epsilon_2(234) = (\epsilon_2 \cup \epsilon_2)(01234)$.}
\end{center}
\end{figure}

Let us study the 15j symbol of this 2-category. This is the quantity we will multiply 4-simplex-by-4-simplex along a triangulation of $X$ to obtain the partition function of the $G$-crossed 2-Ising theory. As we see from Fig. 5, the 15j symbol is the exponential of 
$$\hat \alpha_4 = \nu + \frac{1}{2} \epsilon_2 \cup_1 \rho + \frac{1}{2} \epsilon_2 \cup \epsilon_2.$$
This is analogous to Eq (3.12) in \cite{BGK}. Evaluating the partition function on the boundary of a 5-ball should give $1$, which is equivalent to the condition  $\delta \hat \alpha_4 = 0$. Since $\delta\epsilon_2 = \rho(A)$, this  is equivalent to the Gu-Wen equation
$$
\delta \nu = \frac{1}{2}\rho \cup_1 \rho.
$$

Now we can write the partition function in a fixed $G$ background $A \in Z^1(X,G)$ as (up to positive multiplicative factors)
$$
Z(X,A) \simeq \sum_{\epsilon_2 \in C^2(X,\ZZ_2)|\delta\epsilon_2 = \rho(A)} \exp{2\pi i \int_X \left(\nu(A) +\frac{1}{2} \epsilon_2 \cup_1 \rho(A) +\frac{1}{2} \epsilon_2 \cup \epsilon_2\right)}.
$$

Consider now coupling this theory to a background 3-form gauge field  $C \in Z^3(X,\ZZ_2)$. This is achieved by replacing the constraint (\ref{eps2rhoconstraint}) with
\begin{equation}
\delta\eps_2=\rho(A)+C.
\end{equation}
This ensures symmetry under the 2-form $\ZZ_2$ gauge symmetry $\eps_2\mapsto\eps_2+\beta$, $C\mapsto C+\delta\beta$, where $\beta\in C^2(X,\ZZ_2)$. Thus we must merely replace $\rho(A)$ with $\rho(A)+C$. The partition function is thus
$$
Z(X,A,C) \simeq \sum_{\substack{\epsilon_2 \in C^2(X,\ZZ_2) \\\delta\epsilon_2 = C + \rho(A)}}\exp{2\pi i \int_X \left(\nu(A) +\frac{1}{2} \epsilon_2 \cup_1 C +\frac{1}{2} \epsilon_2 \cup_1 \rho(A) +\frac{1}{2} \epsilon_2 \cup \epsilon_2\right)}.$$
Now consider the effect of the gauge transformation $C\mapsto C+\delta\beta$. Making a change of variables $\eps_2\mapsto\eps_2+\beta$, we find after some work:
$$
Z(X,A,C+\delta\beta)= Z(X,A,C) \exp{2\pi i \int_X \left(\frac12(C+\rho(A))\cup_2 \delta\beta+ \frac{1}{2} \beta \cup \beta+\frac{1}{2}\beta \cup_1 \delta\beta\right)}.
$$
Observe the appearance of the first descendant of $Sq^2 C$ (see Appendix \ref{descent})
evaluated at the value of the 3-form gauge field $C+\rho(A)$. This is almost the expected transformation law for the partition function, except that we expect $C$, not $C+\rho(A)$. This is easily fixed by multiplying the partition function $Z(X,A,C)$ by an additional factor
$$
\exp{2\pi i\int_X \rho(A)\cup_2 C}. 
$$
This is a non-minimal contact-term coupling between $C$ and $A$ ensuring that  the model has the proper anomaly for the 2-form symmetry to be a bosonic shadow of a fermionic phase. In fact, it also shows that the full $E$ symmetry of this theory has the $\tilde Sq^2_\pm (C,B)$ symmetry, since by our results in Appendix \ref{Eanomalies}, this is determined once one knows the anomaly for the $C$ part only, though we cannot decide whether to take the $+$ or $-$ extension. We leave the explicit construction of state sums for more general 3d FSREs to future work.

\section{Fermionic string phases}
As discussed in Appendix \ref{Eanomalies}, the two possible anomalies for a bosonic shadow of a fermionic theory are $\tilde Sq^2_\pm (C,B)$ which differ by
\begin{equation}\label{BSqBanomaly}
\frac12\int_P B\cup Sq^1B,\quad B\in Z^2(P,\ZZ_2).
\end{equation}
In this section we would like to investigate the physics of this term alone. That is, we consider a bosonic theory with a 1-form $\ZZ_2$ symmetry and an anomaly given by (\ref{BSqBanomaly}). Since the 3-form gauge field $C$ does not enter the anomaly, it is irrelevant whether the 2-form $\ZZ_2$ symmetry is present or not. If it is present, one can gauge it without introducing the spin structure (since the 2-form $\ZZ_2$ symmetry is nonanomalous now) and reduce to the case when it is absent.

Since the anomaly (\ref{sqtilde}) is trivialized by the spin structure, and (\ref{BSqBanomaly}) is twice (\ref{sqtilde}), this means that the latter anomaly is also trivialized by the spin structure. To see this more directly, we use the following  identities in $H^5(P,U(1))$, where $P$ is any closed oriented 5-manifold  (see Appendix \ref{oneformanomalies}):
$$
\frac12 B\cup Sq^1B=\frac12 Sq^2 Sq^1 B=\frac12 [w_2(P)]\cup Sq^1 B. 
$$
For a closed spin 5-manifold $P$, $[w_2(P)]=0$, so the anomaly is trivial. But there are no fermionic particles, because the 2-form symmetry, even if present, is not anomalous, and gauging it merely  leads to a condensation of bosons (the worldlines of the $C$-field). It seems that we have a violation of the
\[{\rm spin}\implies{\rm statistics}\]
relation unless one considers also string statistics on the right hand side. 

In fact, it appears possible to fermionize the theory with something less restrictive than a spin structure: a $w_3$-structure \cite{T}. Just like a spin structure can be thought of as a trivialization of $w_2$, a $w_3$-structure on an oriented $n$-manifold $Z$ is a 2-cocycle $\Gamma\in C^2(Z,\ZZ_2)$ such that $\delta\Gamma=w_3$, defined up to exact 2-cocycles. Clearly, any two $w_3$ structures differ by an element of $H^2(Z,\ZZ_2)$, so the set of $w_3$ structures can be identified with $H^2(Z,\ZZ_2)$, but not canonically. To see the relevance of $w_3$-structures, we note that
$w_3=Sq^1 w_2$, hence
$$
[w_2]\cup Sq^1 B = [w_3]\cup B. 
$$
Hence the anomaly is trivial on a closed orientable 5-manifold $P$ satisfying $[w_3(P)]=0$. On a 5-manifold with a boundary $X$, we need a trivialization $\Gamma$ of $w_3(X)$ to define a counterterm $\int_X \Gamma\cup B$ which cancels the anomaly. 

A model which depends on a $w_3$-structure but does not have fermions evades the contradiction with the spin-statistics relation. But it does not correspond to a normal bosonic phase either. In the remainder of this section we make a few remarks about such unusual phases.

First, although not every closed oriented 4-manifold is spin (a counter-example being $\CC\PP^2$), every closed oriented 4-manifold admits a $w_3$ structure. This can be easily shown using $w_3=Sq^1w_2$ and properties of Steenrod squares. 

Second, gauging a 1-form $\ZZ_2$-symmetry means proliferating strings. Their worldsheets are Poincar\'{e}-dual to $B\in Z^2(X,\ZZ_2)$. The anomalous nature of the 1-form symmetry means that these strings need a $w_3$-structure for their definition. Such strings were discussed recently in a somewhat different context by one of us \cite{T} and were dubbed fermionic strings. Their normal bundle is framed, and the wavefunction is multiplied by $-1$ when the framing is twisted by one unit. Thus phases requiring $w_3$ structure may be called fermionic string phases. 

A simple way to construct a fermionic string phase is to start with a bosonic model with a 2-form $\ZZ_2$ symmetry and anomaly (\ref{GuWenanomaly}) and set $C=Sq^1B$. This means that we are embedding the 1-form $\ZZ_2$ symmetry into the 2-form $\ZZ_2$ symmetry group and then gauge the 1-form symmetry. 
The resulting theory clearly has no $\ZZ_2$-grading on its Hilbert space, because $\int_Y Sq^1B$ vanishes for any oriented 3-manifold $Y$. This means that in general fermionic string phases do not have a conserved $\ZZ_2$-valued charge analogous to $(-1)^F$. It also illustrates that ordinary fermionic phases do not really come in two types corresponding to the $\pm$ in $\tilde Sq^2_\pm$ because we can flip the sign by a redefinition of the symmetry operators corresponding to the shift $C \mapsto C + Sq^1 B$.

A further insight is obtained by noticing that while the homology 2-cycle dual to $B$ represents the string worldsheet $\Sigma$, the homology 1-cycle dual to $Sq^1B$ can be thought of as the 1-cycle on $\Sigma$ which is dual (in the 2d sense) to the 1st Stiefel-Whitney class of the normal bundle of $\Sigma$. Note that while the 4-manifold $X$ is assumed to be oriented, $\Sigma$ need not be orientable. Since we may think of $C$ as $Sq^1B$, we conclude that fermionic string phases have fermion worldlines confined to fermionic string worldsheets. 

Finally, let us give a couple of examples of bosonic shadows of fermionic string phases. First, as remarked above we can take a shadow of any standard fermionic phase (with a 2-form $\ZZ_2$ symmetry only) and embed the 1-form $\ZZ_2$ symmetry into the 2-form $\ZZ_2$ symmetry. For example, we can take the model (\ref{WW}) and consider a global 1-form $\ZZ_2$ symmetry which acts as follows:
\begin{equation}
a\mapsto a,\quad b\mapsto b+\lambda\cup\lambda,\quad \lambda\in Z^1(X,\ZZ_2).
\end{equation}
It is easy to see that the action is invariant for any closed oriented $X$ and any $\lambda$. 

Another way to obtain a shadow of a fermionic string phase is to start with a model with both a 1-form and a 2-form $\ZZ_2$ symmetries and a mixed anomaly
$$
\frac12 \int_P C\cup B
$$
and then set $C=Sq^1 B$. That is, we embed the 1-form $\ZZ_2$ symmetry into a product of 1-form and 2-form $\ZZ_2$ symmetries in a nonstandard way. As a simple example, consider the $\ZZ_2$ gauge theory in 3+1d with an action
\begin{equation}\label{4dZ2theory}
\frac12 \int_X b\cup \delta a,\quad b\in C^2(X,\ZZ_2),\ a\in C^1(X,\ZZ_2). 
\end{equation}
One can get the desired anomaly (\ref{BSqBanomaly}) by considering the following action of a global 1-form $\ZZ_2$ symmetry:
$$
a\mapsto a+\lambda,\quad b\mapsto b+\lambda\cup\lambda,\quad\lambda\in Z^1(X,\ZZ_2).
$$
Gauging this symmetry means proliferating the strings and the particles of the $\ZZ_2$ gauge theory, but with particles confined to the string worldsheets in a particular way. Since the particles are not local with respect to the strings, one is forced to choose a $w_3$-structure on $X$ to make the result well-defined.

\section{Concluding remarks}

We have argued that every 3d fermionic model has a bosonic shadow which has a certain 3-group symmetry $E$ with an anomaly. Further, we argued that 3d FSREs with a finite unitary symmetry $G$ are classified by triples $(\nu,\rho,\sigma)$ satisfying certain rather complicated equations generalizing the Gu-Wen supercohomology. We proposed that bosonic shadows of all such models are $G$-equivariant versions of a certain 4d TQFT which we called the 2-Ising model. If $\sigma\in H^2(G,\ZZ_2)$ vanishes, we can replace this 4d TQFT with the simplest Crane-Yetter-Walker-Wang model and recover the supercohomology phases. 

Gauging the anomalous 3-group symmetry $E$ is achieved by proliferating fermionic particles and Kitaev strings. It would be interesting to construct explicitly the resulting lattice model. 

Our proposed classification of 3d FSRE phases can be made concrete once we pick a particular symmetry group $G$. Let us give a few examples. If $G=\ZZ_n$ with $n$ odd, both $\rho$ and $\sigma$ vanish, and 3d FSRE phases are classified by the same data as bosonic FSRE phases. If $G=\ZZ_n$ with $n$ even, we only get Gu-Wen supercohomology phases, because the parameter $\sigma$ vanishes. Indeed, while $H^2(\ZZ_n,\ZZ_2)\simeq \ZZ_2$ if $n$ is even, the generator of this $\ZZ_2$ does not square to zero (in fact, it generates a polyominal ring inside $H^{\bullet}(\ZZ_n,\ZZ_2)$ \cite{Evens}). Hence the equation $\delta\rho=\sigma\cup\sigma$ has solutions only if $[\sigma]=0$. Similarly, if $G$ is a product of several copies of $\ZZ_2$, we only get supercohomology phases, since the cohomology ring of $G$ with $\ZZ_2$ coefficients is a polynomial ring \cite{Evens}, and any element of $H^2(G,\ZZ_2)$ which squares to zero must be trivial. The simplest example where there are phases which are not supercohomology phases is $G=\ZZ_4\times\ZZ_2$. $H^{\bullet}(G,\ZZ_2)$ is generated by two elements $x,y$ of degree $1$ and an element $w$ of degree $2$. The only relation is $x^2=0$. Thus $H^2(G,\ZZ_2)$ has a unique nontrivial nilpotent element $xy$. If we set $\sigma=xy$, we get 3d FSRE phases where the  zipper $\zp(g_4,g_2)$, where $g_4$ and $g_2$ is a Kitaev string, while all other zippers are ``trivial''. Overall, for $G=\ZZ_4\times\ZZ_2$ we get four supercohomology phases (including the trivial phase), and four non-supercohomology phases. 

We also found a new class of phases which are neither bosonic, nor fermionic, in that they have ``fermionic strings'' but no fermionic particles. Their partition function depends on a $w_3$-structure on the 4-manifold. It would be very interesting to explore the physics of these new phases. 

Our discussion was not as systematic as that of \cite{BGK} because we lack an algebraic description of completely general unitary 4d TQFT. In particular, while we outlined the structure of various monoidal 2-categories relevant to us, we did not describe all the data which enter into a definition of these objects. We hope to return to this issue in the future. 

Our results suggest that bosonization in higher dimensions will get progressively more complicated as the dimension increases. This complexity reflects the topological complexity of the spin-bordism spectrum. For example, in four spatial dimensions presumably one would have to deal with a symmetry 4-group which involves 3-form, 2-form and 1-form symmetries. Bosonic shadows of Gu-Wen supercohomology phases would be quite special since they only possess 3-form symmetries. 


\appendix

\section{Steenrod squares and Stiefel-Whitney classes}\label{steenrodreview}

We review here some definitions and results from \cite{Steenrod_higher} and \cite{MilnorStasheff}.

In this paper we mostly work with simplicial cochains of a triangulated manifold $X$ with values in $\ZZ_2$. We assume a local order on vertices of the triangulation. There is a coboundary operation $\delta: C^p(X,\ZZ_2)\ra C^{p+1}(X,\ZZ_2)$ satisfying $\delta^2=0$. As usual, a cochain annihilated by $\delta$ is called a cocycle, and the space of $p$-cocycles is denoted $Z^p(X,\ZZ_2)$. The cohomology class of a cocycle $a$ is denoted $[a]$.

There is a well-known product operation
$$
a\cup b\in C^{p+q}(X,\ZZ_2),\quad a\in C^p(X,\ZZ_2),\ b\in C^q(X,\ZZ_2).
$$
It is bilinear and associative and satisfies the Leibniz rule
$$
\delta(a\cup b)=\delta a\cup b+a\cup\delta b.
$$
The cup product is not (super)commutative on the level of chains, rather one has
$$
a\cup b+b\cup a=a\cup_1\delta b+\delta a\cup_1 b+\delta(a\cup_1 b),
$$
where the new product $\cup_1$ has degree $-1$:
$$
a\cup_1 b\in C^{p+q-1}(X,\ZZ_2),\quad a\in C^p(X,\ZZ_2),\ b\in C^q(X,\ZZ_2).
$$
Note that the cup product is commutative on the level of cohomology classes, i.e. $[a]\cup [b]=[b]\cup[a]$.

The $\cup_1$ product is not commutative either, rather one has
$$
a\cup_1 b+b\cup_1 a=a\cup_2\delta b+\delta a\cup_2 b+\delta(a\cup_2 b),
$$
where yet another product $\cup_2$ appears, etc. 

One defines an operation $Sq^q: H^p(X,\ZZ_2)\ra H^{p+q}(X,\ZZ_2)$, $p\geq q$, by the following  formula on the cochain level:
$$
Sq^q a=a\cup_{p-q} a,\quad a\in Z^p(X,\ZZ_2). 
$$
Despite appearances, this operation is linear on the level of cohomology classes, i.e. $Sq^1[a+b]=Sq^1[a]+Sq^1[b].$ We note that $Sq^1$ is a differential, i.e. 
$$
Sq^1 ([a]\cup [b])=Sq^1 [a]\cup [b]+ [a]\cup Sq^1 [b].
$$
Note also that if $[a]\in H^p(X,\ZZ_2)$, then $Sq^p [a]=[a]\cup [a]$. In particular, for any $[a]\in H^1(X,\ZZ_2)$ one has $[a]\cup[a]=Sq^1[a]$. 

On an $n$-manifold $X$ we have Stiefel-Whitney classes $w_k\in H^k(X,\ZZ_2)$, $k=0,\ldots,n$. The class $w_1$ is an obstruction to orientability. If $w_1$ vanishes, then the class $w_2$ is an obstruction to having a spin structure. 
These classes satisfy a number of relations. In particular, $w_3=Sq^1 w_2$ for all $n$. There also relations which depend on $n$. For example, for $n=2$ we have $w_2=w_1^2$, so any orientable 2-manifold admits a spin structure. 

On a closed $n$-manifold $X$ we also have the Wu formula:
$$
Sq^{n-p} [a]=v_{n-p}\cup [a],\quad [a]\in H^p(X,\ZZ_2),
$$
where $v_{n-p}\in H^{n-p}(X,\ZZ_2)$ is a certain polynomial in Stiefel-Whitney classes independent of $X$. It is known as the Wu class. The lowest Wu classes are $v_1=w_1$, $v_2=w_1^2+w_2$, $v_3=w_1 w_2$. 

The Wu formula has many useful consequences. For example, it implies that on any orientable $n$-manifold one has $Sq^1[a]=0$ for any $a\in H^{n-1}(X,\ZZ_2)$. In particular, on a Riemann surface $X$ the square of every element of $H^1(X,\ZZ_2)$ vanishes.  Another consequence is that on a spin 4-manifold $X$ the square of every element in $H^2(X,\ZZ_2)$ vanishes.

\section{Anomaly descendants}\label{descent}

The 't Hooft anomalies reveal themselves in how the symmetry algebra is realized projectively by unitary operators on the Hilbert space. For an anomaly $\omega$ coming from group cohomology, one can compute the so-called descendants to find the class $c_\omega$ of the projective action. Conversely, a system with a projective symmetry in class $c_\omega$ has the anomaly $\omega$. In this section, we discuss the descent procedure for anomalies of 1-form and 2-form $\ZZ_2$ symmetries relevant for fermionization. Unfortunately the calculation of descendants of the $E$ symmetry anomaly is beyond the scope of this paper.

\subsection{1-form $\ZZ_2$ symmetry in 2+1d}

As a warm-up, we consider the 1-form $\ZZ_2$ symmetry in 2+1d with the anomaly 
$$
\omega(B) = \frac{1}{2} Sq^2 B = \frac{1}{2} B^2 \in H^4(K(\ZZ_2,2),U(1)).
$$
This cohomology class also defines an effective action for a bosonic SPT in 3+1d protected by the 1-form $\ZZ_2$ symmetry. We can use it to compute the SPT ground state on a closed oriented 3-manifold $X$. We consider the path integral on the cone with base $X$, denoted $CX$. $CX$ can be made from the cylinder $X \times [0,1]$ by collapsing $X \times \{0\}$ to a point. From this description one sees that a 2-form gauge field on the cone $B: CX \to K(\ZZ_2,2)=K$ is the same thing as a homotopy from the trivial gauge field on $X\times\{0\}$ to some other gauge field on $X\times\{1\}$. This is of course the same thing as a gauge transformation on $X$ and is parametrized by a $\ZZ_2$ 1-cochain $\lambda \in C^1(X,\ZZ_2)$.

Computing the sum over all these $\lambda$ where we remember the (ungauged) boundary condition on $X$ we obtain the state
$$
|\omega\rangle = \sum_\lambda \exp (i\pi \int_{CX} \delta\lambda\cup \delta\lambda) |\lambda\rangle.
$$
We can rewrite this state as 
$$
|\omega\rangle = \sum_\lambda \exp( i\pi \int_X \lambda\cup \delta\lambda) |\lambda\rangle.
$$
This expression makes it clear that $|\omega\rangle$ is short-range entangled, since it is produced from a product state $\sum_\lambda |\lambda\rangle$ by time $2\pi$ evolution of a Hamiltonian defined by
$$
\omega_1(0,\lambda) = \frac{1}{2} \lambda\cup \delta\lambda
$$
This function is called the first descendant of $\omega(B) = B^2/2$. It also determines the variation of the partition function of the 2+1d theory under the 1-form gauge symmetry transformation. 

Now we consider a global symmetry transformation $\lambda \mapsto \lambda + \beta$, where $\beta$ is a $\ZZ_2$ 1-cocycle, $\beta\in Z^1(X,\ZZ_2)$. We find a variation in the exponent
$$
\omega_1(0,\lambda+\beta) - \omega_1(0,\lambda) = \frac{1}{2}\beta\cup \delta\lambda = \frac{1}{2} \delta(\beta\cup\lambda)
$$
which defines the second descendant
$$
\omega_2(0,\lambda,\beta) = \frac{1}{2} \beta\cup\lambda.
$$
This means that for closed $X$ the SPT ground state is invariant under global  1-form symmetry, but when $\partial X$ is nonempty, it is invariant only up to a boundary term:
$$
|\omega\rangle\mapsto \sum_\lambda \exp(i\pi \int_X \lambda\cup d\lambda + i\pi \int_{\partial X} \beta\cup\lambda)|\lambda\rangle.
$$
It also tells us whether the global 1-form symmetry acts projectively on the Hilbert space of the 2+1d theory. That is, whether transforming by $\beta_1+\beta_2$ is any different than transforming by $\beta_1$ followed by $\beta_2$. This is measured by
$$
\omega_2(0,0,\beta_1+\beta_2) - \omega_2(0,\beta_1,\beta_2) - \omega_2(0,\beta_1,0) = \frac{1}{2} \beta_1 \cup \beta_2,
$$
so indeed we do have a projective symmetry action measured by a bilinear form on the symmetry 2-group $B\ZZ_2$. As we have seen above, such cocycles are trivialized by quadratic refinements of the form, which in this dimension we may obtain from a spin structure.

\subsection{2-form $\ZZ_2$ symmetry in 3+1d}

Next we consider 2-form $\ZZ_2$ symmetry in 3+1d with an anomaly
$$
\omega(C)=\frac12 Sq^2C=\frac12 C\cup_1C\in H^5(K(\ZZ_2,3),U(1)).
$$
where $[C]$ is the generator of $H^3(K(\ZZ_2,3),U(1))\simeq\ZZ_2$. It can also be regarded as an effective action of a 4+1d SPT with a 2-form $\ZZ_2$ symmetry. As before, on a 4-manifold $X$ we obtain an SPT ground state
$$
|\omega\rangle = \sum_\Lambda \exp(i\pi \int_{CX} \delta\Lambda \cup_1 \delta\Lambda)|\Lambda\rangle,
$$
where $\Lambda \in C^2(X,\ZZ_2)$ is a $\ZZ_2$-valued 2-cochain parametrizing a gauge transformation of $C$. We compute
$$
\delta\Lambda \cup_1 \delta\Lambda = \delta(\Lambda \cup_1 \delta\Lambda + \Lambda\cup\Lambda) \mod 2.
$$
So we can rewrite
$$
|\omega\rangle = \sum_\Lambda \exp(2 i\pi \int_{X}\omega_1(0,\Lambda))|\Lambda\rangle
$$
using the first descendant
$$
\omega_1(0,\Lambda) = \frac{1}{2}(\Lambda \cup_1 \delta\Lambda + \Lambda\cup\Lambda).
$$
Now we compute the transformation of this state under a global symmetry $\Lambda \mapsto \Lambda + \beta$ for some $\ZZ_2$-valued 2-cocycle $\beta \in Z^2(X,\ZZ_2)$. We find 
$$
|\omega\rangle \mapsto \exp(i\pi \int_X \beta^2)\sum_\Lambda \exp(i\pi \int_{X}\left(\Lambda \cup_1 \delta\Lambda + \Lambda\cup\Lambda\right) + i\pi \int_{\partial X} \beta \cup_1 \Lambda)|\Lambda\rangle.
$$
This variation involves the boundary variation we expected from the previous calculation but also a new ingredient: a prefactor
$$
\exp(i\pi \int_X \beta\cup\beta).
$$
This factor is not a boundary term for a general $X$ and $\beta$. For example, when $X = \mathbb{CP}^2$ and $\beta$ represents the unique nonzero degree-2 class, this prefactor is $-1$. Such a symmetry transformation multiplies the ground state by $-1$. Thus we are dealing with an SPT phase only if we restrict to those $\beta$ for which $[Sq^2\beta]=[\beta\cup\beta]=0$. Alternatively, we can restrict $X$ to be a spin 4-manifold.

After this is done, we can extract the second descendant which measures to what extent the action of the global 2-form symmetry on the Hilbert space of the 3+1d theory is projective:
$$
\omega_2(0,\beta',\beta) = \frac{1}{2}\beta\cup_1 \beta'.
$$
An interesting feature about this term is that it is not invariant under 2-gauge transformations $\beta \mapsto \beta + \delta \lambda$, where $\lambda \in C^1(X,\ZZ_2)$. Indeed, it has a variation
$$
\delta \omega_2(0,\beta',\beta) = \frac{1}{2} \delta\lambda \cup_1 \beta'
$$
which looks like it could be exact but unfortunately is not. This means that the symmetry action on the boundary is only defined for cocycles $\beta \in Z^2(\partial X,\ZZ_2)$ and not cohomology classes. Likewise, the bilinear form
$$
\omega_2(0,0,\beta_1+\beta_2) - \omega_2(0,\beta_1,\beta_2)-\omega_2(0,0,\beta_1) = \frac{1}{2} \beta_1 \cup_1 \beta_2
$$
is only well-defined on $Z^2(\partial X,\ZZ_2)$.

\section{The function $Q_\Eta(B)$}\label{QEta}

We summarize here the definition and properties of the function 
$$
Q_\Eta(B): Z^2(Y,\ZZ_2)\ra \ZZ_2.
$$
Here $Y$ is a closed oriented 3-manifold equipped with a triangulation and a  branching structure, and $\Eta$ is a spin structure on $Y$. 

Given an oriented 3-manifold $Y$, there exists an oriented 4-manifold $X$ with boundary $Y$ such that any $B\in Z^2(Y,\ZZ_2)$ extends to a 2-cocycle $B_X$ on $X$. In other words, the restriction map $H^2(X,\ZZ_2)\ra H^2(Y,\ZZ_2)$ is surjective. Given such $X$, we may consider the relative Stiefel-Whitney class $w_2(X,Y,\Eta)\in H^2(X,Y,\ZZ_2)$ which measures the obstruction to extending $\Eta$ to a spin structure on $X$. Recall also that the cup product makes $H^{\bullet} (X,Y,\ZZ_2)$ into a module over the algebra $H^{\bullet}(X,Y)$, and that the fundamental homology class $[X]$ takes values in $H_4(X,Y,\ZZ_2)$. Thus it makes sense to consider the expression
\begin{equation}\label{w2Bevaluation}
[B_X]\cup w_2(X,Y,\Eta)\cap [X].
\end{equation}
We write it somewhat schematically as
$$
\int_X w_2\cup B_X+\int_Y \Eta\cup B,
$$
to indicate that when the spin structure $\Eta$ is shifted by $\alpha\in Z^1(Y,\ZZ_2)$, the quantity (\ref{w2Bevaluation}) shifts by 
$$
\int_Y\alpha\cup B.
$$
In other words, the set of spin structures on $Y$ is an affine space over the vector space $H^1(Y,\ZZ_2)$, and the quantity (\ref{w2Bevaluation}) is an affine linear function on it.

The quantity (\ref{w2Bevaluation}) depends on the cohomology class $[B_X]\in H^2(X,\ZZ_2)$, but not on the concrete representative. But it also depends on the choice of $X$ as well as the choice of the extension of $B$ from $Y$ to $X$. Given two such choices, $(X,B_X)$ and $X',B_{X'}$, the difference between the corresponding expressions is
$$
\int_T w_2\cup B_T
$$
where the 4-manifold $T$ obtained by gluing $X$ and $X'$ along $Y$ is closed, and $B_T$ restricts to $B_X$ on $X$ and $B_{X'}$ on $X'$. But this is the same as
$$
\int_T B_T\cup B_T.
$$

This implies that the expression
$$
Q_\Eta(B)=\int_X B_X\cup B_X + \int_X w_2(X,Y,\Eta)+\int_Y \Eta\cup B
$$
does not depend either of the choice of extension of $B$ from $Y$ to $X$, nor on the choice of $X$. On the other hand, since $B_X$ is an absolute 2-cocycle, it does depend on the choice of $B$ within its cohomology class in $H^2(Y,\ZZ_2)$. It is easy to see that
$$
Q_\Eta(B+\delta\lambda)=\int_Y \left(\lambda\cup\delta\lambda+\delta\lambda\cup_1 B\right).
$$
It is also easy to see that 
$$
Q_{\Eta+\alpha}(B)=Q_\Eta(B)+\int_Y \alpha\cup B,\quad \forall \alpha\in Z^1(Y,\ZZ_2).
$$

Now suppose the spin structure $\Eta$ extends to $X$. Then $w_2(X,Y,\Eta)$ vanishes, and therefore we get
$$
Q_\Eta(B)=\int_X B_X\cup B_X.
$$
This property is used in section 4 to argue the conservation of fermion number.

Conversely, for a fixed triangulation and branching structure, the function $Q_\Eta$ completely determines the equivalence class of $\Eta$. Indeed, given any two spin structures $\Eta$ and $\Eta'$, the difference $(Q_\Eta-Q_{\Eta'})(B)$ is linear and depends only on the cohomology class of $B$ and thus must have the form $\int_Y \alpha\cup B$ for some $\alpha\in Z^1(Y,\ZZ_2)$. On the other hand, this difference is equal to $\int_Y (\Eta-\Eta')\cup B$. By Poincar\'{e} duality, $Q_\Eta(B)=Q_{\Eta'}(B)$ for all $B$ implies that $\Eta-\Eta'$ is exact, which means that $\Eta\sim\Eta'$.

\section{'t Hooft anomalies for a 1-form $\ZZ_2$ symmetry}\label{oneformanomalies}

In this section we classify possible anomalies for a 1-form $\ZZ_2$ symmetry in 3+1d, both for bosonic and fermionic theories, assuming the space-time symmetry is orientable (i.e. ignoring time-reversal symmetries, if any). 

In the bosonic case, we need to compute the oriented cobordism group $\Omega^5_{SO}(K(\ZZ_2,2),U(1)).$ It is the Pontryagin-dual of the oriented bordism group $\Omega_5^{SO}(K(\ZZ_2,2),\ZZ)$. Physically, these classify possible 5d topological actions built out of a 2-form gauge field $B\in Z^2(P,\ZZ_2)$, where $P$ is a closed oriented 5-manifold. All such topological terms will be integrals of densities made out of $B$ and certain characteristic classes of the tangent bundle of $P$, namely the Stiefel-Whitney classes and the Pontryagin classes. Actually, since Pontryagin classes modulo 2 can be expressed through Stiefel-Whitney classes, it is sufficient to consider the latter. An obvious approach is to construct elements in $H^5(P,\ZZ_2)$ and then embed them into $H^5(P,U(1))$ using the embedding $\ZZ_2\ra U(1)$, keeping in mind that distinct elements of $H^5(P,\ZZ_2)$ can become identical elements of $H^5(P,U(1))$. 

Let us write down candidate independent terms in $H^5(P,\ZZ_2)$. Orientability implies $w_1(P)=0$ and it follows $Sq^1 x=0$ for any $x\in H^4(P,\ZZ_2)$ \cite{MilnorStasheff}, so we have four candidates:
$$
B Sq^1B,\quad Sq^2 Sq^1 B,\quad w_2 Sq^1B,\quad w_3 B.
$$
The 2nd and the 3rd are actually the same thanks to the Wu formula \cite{MilnorStasheff}. Further, since $Sq^1$ satisfies the Leibniz rule w. r. to the cup product, and $w_3=Sq^1 w_2$, the 4th one is the same as the 3rd one.
Thus we are left with only two independent elements of $H^5(P,\ZZ_2)$. Now we must map these classes to $H^5(P,U(1))$. In fact, we will find they map to the same (nonzero) element. To see this, one needs to use the long exact sequence 
$$
\ldots\ra H^4(K,U(1))\ra H^4(K,U(1))\ra H^5(K,\ZZ_2)\ra H^5(K,U(1))\ra \ldots
$$
where we have introduced the short-hand $K=K(\ZZ_2,2)$. The 1st map is multiplication by $2$, and the 2nd map is the Bockstein homomorphism associated to the short exact sequence $\ZZ_2\ra U(1)\ra U(1)$. We are interested in the image of the Bockstein homomorphism. The group $H^4(K(\ZZ_2,2),U(1))$ is isomorphic to $\ZZ_4$ and is generated by $1/4$ times the Pontryagin square of $[B]\in H^2(K(\ZZ_2,2),\ZZ_2)$, for which a representative may be written
\begin{equation}\label{Pontryaginsq}
\frac{1}{4} (\tB\cup \tB+\delta\tB\cup_1 \tB),
\end{equation}
where $\tB\in C^2(K(\ZZ_2,2),\ZZ)$ is an integral lift of $B$. Using the defining property of $\cup_1$, we find that the Bockstein of (\ref{Pontryaginsq}) is
$$
\frac{1}{4} \tB\cup \delta\tB+\frac{1}{8} \delta\tB\cup_1\delta\tB,
$$
which is a cocycle formula for $BSq^1B+Sq^2Sq^1B$ in $Z^5(K,\ZZ_2)$. By exactness, it follows that this difference maps to zero in $H^5(K,U(1))$.

To verify that this exhausts all possible topological actions, one can use the Atiyah-Hirzebruch spectral sequence computing oriented bordism groups starting from the homology groups $H_p(K,\Omega_q^{SO}(\star))$. The integral homology groups of $K=K(\ZZ_2,2)$ are known, and the nonzero ones up to degree $5$ are
$$
H_0(K)=\ZZ,\quad H_2(K)=\ZZ_2, \quad H_4(K)=\ZZ_4,\quad H_5(K)=\ZZ_2.
$$
The first term defines a purely gravitational anomaly $w_2w_3$ which splits off from the anomaly group. Then the spectral sequence implies that the map $\Omega_5^{SO}(K)\ra H_5(K) \oplus w_2 w_3$ is an isomorphism. Hence the group of gauge anomalies $\tilde\Omega^5_{SO}(K,U(1))=\ZZ_2$, and the nontrivial anomaly action can be written as
\begin{equation}\label{bosonic5d}
\frac12\int_P B Sq^1B=\frac12\int_P Sq^2Sq^1B=\frac12 \int_P w_2 Sq^1B. 
\end{equation}

In the fermionic case, we need to compute $\Omega^5_{Spin}(K,U(1))$. Note that since $w_2(P)=0$ for a closed orientable spin manifold $P$, the bosonic action (\ref{bosonic5d}) becomes trivial for such $P$. That is, the image of the map $\Omega^5_{SO}(K,U(1))\ra \Omega^5_{Spin}(K,U(1))$ is trivial. However, there can also be elements of $\Omega^5_{Spin}(K,U(1))$ which do not come from $\Omega^5_{SO}(K,U(1))$. These topological terms use the spin structure in a key way and won't be just integrals of characteristic classes. Looking at the Atiyah-Hirzebruch spectral sequence, we see that the only such element arises from the $E_2$ term $H^4(K,\ZZ_2)$. This cohomology group is $\ZZ_2$ and is generated by $[B\cup B]=[Sq^2B]$. The corresponding spin-topological action is evaluated as follows: we take the homology class in $H_1(P,\ZZ_2)$ which is Poincar\'{e}-dual to $[B\cup B]$ and pick a closed 1d submanifold $\gamma$ which realizes it. Then we restrict the spin structure of $P$ to $\gamma$ and evaluate the corresponding holonomy. Thus $\Omega^5_{Spin}(K,U(1))=\ZZ_2$. 

Note that if $B$ satisfies the constraint $[B\cup B]=0$, then the corresponding 5d spin-topological action is zero. Thus the fermionic anomaly is necessarily trivial for such $B$.

\section{'t Hooft Anomalies for a 2-form $\ZZ_2$ symmetry}\label{twoformanomalies}

In this section we wish to discuss possible 't Hooft anomalies for bosonic and fermion systems with 2-form $\ZZ_2$ symmetry. The calculations are much the same as the previous section, except where now the classifying space $K=K(\ZZ_2,3)$, an Eilenberg-Maclane space with only nonzero homotopy group $\pi_3 = \ZZ_2$. For bosonic systems, the possible anomalies are
\[Sq^2 C, \quad w_2 C, \quad w_2 w_3.\]
The first two are actually equal thanks to the Wu formula, while the third does not involve the gauge field $C \in Z^3(P,\ZZ_2)$ and so describes a purely gravitational anomaly. So as before, we find that (3-)group cohomology describes all the gauge anomalies.

When we consider these terms on a closed spin 5-manifold $P$, they all vanish because $w_2 = 0$. For new anomalies we look to $C \in H^3(K,\ZZ_2)$ and $Sq^1 C \in H^4(K,\ZZ_2)$ in the Atiyah-Hirzebruch spectral sequence. The first does not survive the $d_2$ differential since $d_2 C = Sq^2 C \neq 0$. The second, however, defines a topological term which measures the holonomy of the spin structure along the curve Poincar\'e dual to $Sq^1 C$.

\section{'t Hooft Anomalies for $E$ symmetry}\label{Eanomalies}

We have discussed the appearance of the 3-group symmetry $E$, whose elements are pairs $(\beta,\alpha) \in Z^2(X,\ZZ_2) \times Z^1(X,\ZZ_2)$. The group law is not the product, but has a twist
$$
(\beta_1,\alpha_1) \circ (\beta_2,\alpha_2) = (\beta_1 + \beta_2 + \alpha_1 \cup \alpha_2,\alpha_1 + \alpha_2).
$$
In this section we discuss possible anomalies for such а symmetrы in bosonic and fermionic systems. Like the $\ZZ_2$ 1-form symmetry considered in the previous appendix, $E$ has a classifying space denoted $BE$ with $\pi_2 = \ZZ_2$, $\pi_3 = \ZZ_2$ and the Postnikov class $Sq^2$, reflecting the twisted group law. This means that one can think of a map $P \to BE$ as a pair $(B,C) \in Z^2(P,\ZZ_2) \times C^3(P,\ZZ_2)$ satisfying
\begin{equation}\label{apostnikov}
dC = B\cup B.
\end{equation}
From this one sees that there is a map $BE \to K(\ZZ_2,2)=K$ by forgetting $C$. This map is a fibration with fiber $K(\ZZ_2,3)=L$. This fibration is very useful for computing the cohomology of $BE$. For instance, to compute $H^5(BE,U(1)) \simeq H^6(BE,\ZZ)$, a first approximation to the bosonic anomaly group $\Omega^5_{SO}(BE,U(1))$, we use the Serre spectral sequence which starts with $E_2^{p,q} = H^p(K,H^q(L))$. The three possible terms are
$$
\frac{1}{2} [C\cup_1 C] \in E_2^{5,0} \quad \frac{1}{2} [B \cup C] \in E_2^{3,2} \quad \frac{1}{2} [B\cup \frac{\delta \hat B}{2}]\in E_2^{0,5},
$$
where $\hat B$ is an integral lift of $B$. The differential in this spectral sequence comes from eq. (\ref{apostnikov}). For example, the differential of the second term above is $\frac{1}{2} B^3$, which is a non-zero class in $H^6(K,U(1))$, so we throw it out. The differentials of the third term are all zero, but it may be a differential of something else. However, there is no candidate in the right degree, so the third term survives to give a nontrivial class in $H^5(BE,U(1))$. The first term is a little complicated, but after some work one finds\footnote{We are grateful to Greg Brumfiel and John Morgan for communicating to us some related results \cite{MorganBrumfiel2}.} that it also survives, and that the corresponding element of $H^5(BE,U(1))$ looks as follows:
\begin{equation}\label{sqtilde}
\tilde Sq^2_\pm(C,B) = \frac{1}{2} C \cup_1 C + \frac{1}{2} C \cup_2 (B^2) \pm \frac{1}{4}\hat B\cup \frac{\delta\hat B}{2} + \frac{1}{2}x(B).
\end{equation}
Here $x(B)$ is a mod-2 cochain defined by the equation $dx(B) = B^2 \cup_2 B^2 + (B\cup_1 B)^2$. An explicit simplicial expression for it is \cite{Medina}
\begin{equation}\label{xformula}
x(B)(012345)=B(023)B(245)B(012)B(235).
\end{equation}

The expression (\ref{sqtilde}) restricts to $\frac12 Sq^2C$ when one sets $B=0$, i.e. it is a extension of $[\frac12 Sq^2C]\in H^5(L,U(1))$ to the total space $BE$. Note that the extension is not unique, and that the two possible extensions differ by $\frac12 [BSq^1B]$. Note also that
\begin{equation}\label{ssextension}
2[Sq^2_\pm(C,B)] = \frac12\left[ B\cup \frac{\delta \hat B}{2}\right]=\frac12 [BSq^1B].
\end{equation}
This means that the cohomology group $H^5(BE,U(1))$ is isomorphic to $\ZZ_4$, and that it is generated by the cohomology class of Eq.~\eqref{sqtilde}, for either choice of the sign. The ambiguity in the sign is simply the ambiguity in choosing the generator of $\ZZ_4$.

The anomalies for 3+1d bosonic systems with $E$ symmetry are actually classified by the cobordism group $\Omega^5_{SO}(BE,U(1))$. There is another useful spectral sequence for computing this, the Atiyah-Hirzebruch-Serre spectral sequence, which goes from $H^p(K_2,\Omega^q_{SO}(K_3,U(1)))$ to $\Omega^{p+q}_{SO}(BE,U(1))$. We have shown in the above section that the map $H^5(K_3,U(1)) \to \tilde\Omega_{SO}^5(K_3,U(1))$ is an isomorphism and it is not hard to show that $H^q(K_3,U(1)) \to \Omega_{SO}^q(K_3,U(1))$ is also an isomorphism for all $q<5$. Thus, except for the purely gravitational anomaly $w_2 w_3$, all the $E$ anomalies are classified by $H^5(BE,U(1)) = \ZZ_4$. Note that the shift $C \mapsto C + Sq^1 B$ exchanges $\tilde Sq^2_+$ and $\tilde Sq^2_-$, so the difference between the anomalies is not really physical, amounting to a redefinition of the symmetry operators. Further, $[\tilde Sq^2_\pm(Sq^1 B,0)] = [\frac{1}{2}B \frac{\delta \hat B}{2}]$.

Now we want to consider what happens with the map $\Omega^5_{SO}(BE,U(1)) \to \Omega^5_{Spin}(BE,U(1))$. For this we can use naturality of the Atiyah-Hirzebruch-Serre spectral sequence. We know that our nonzero classes in $\Omega^5_{SO}(BE,U(1))$ come from $\frac{1}{2}Sq^2 C \in H^5(K_3,U(1))$ and $\frac{1}{2} B Sq^1 B \in H^5(K_2,U(1))$. Because the first is proportional to $w_2$ and the second to $w_3$, these map to zero on closed spin 5-manifolds, which all have $w_2 = 0$ and $w_3 = Sq^1 w_2 = 0$. This implies that the map $\Omega^5_{SO}(BE,U(1)) \to \Omega^5_{Spin}(BE,U(1))$ sends all the gauge anomalies to zero (and $w_2 w_3$ too, for that matter).

\end{document}